\documentclass[aoas]{imsart}

\RequirePackage[OT1]{fontenc}
\RequirePackage{amsthm,amsmath}
\RequirePackage[numbers]{natbib}
\RequirePackage[colorlinks,citecolor=blue,urlcolor=blue]{hyperref}
\RequirePackage{hypernat}

\usepackage{subfigure,epsfig,floatflt,lscape,graphicx,pdflscape}
\usepackage{color,multirow,bm,enumitem}
\usepackage[normalsize]{caption}
\usepackage[nolists,tablesfirst,nomarkers]{endfloat}

\begin{document}
\begin{frontmatter}

\title{A bivariate space-time downscaler under space and time misalignment}
\runtitle{A bivariate space-time downscaler}

\begin{aug}
\author{\fnms{Veronica J.} \snm{Berrocal}\thanksref{m1}\ead[label=e1]{vjb2@stat.duke.edu}},
\author{\fnms{Alan E.} \snm{Gelfand}\thanksref{m2}\ead[label=e2]{alan@stat.duke.edu}}
\and
\author{\fnms{David M.} \snm{Holland}\thanksref{m3}\ead[label=e3]{Holland.David@epa.gov}}

\runauthor{V. J. Berrocal et al.}

\affiliation{SAMSI\thanksmark{m1}, Duke University\thanksmark{m2} and U.S. Environmental Protection Agency\thanksmark{m3}}

\address{SAMSI \\ 19 T.W. Alexander Drive \\ P.O. Box 14006 \\ Research Triangle Park, NC 27709 \\ USA \\ \printead{e1} \\
\phantom{E-mail:\ }}
\address{Department of Statistical Science\\ Duke University \\ Box 90251 \\ Durham, NC 27708 \\ USA \\ \printead{e2} \\
\phantom{E-mail:\ }}
\address{U.S. Environmental Protection Agency \\ National Exposure Research Laboratory \\ Research Triangle Park, NC 27711 \\ USA \\ \printead{e3} \\
\phantom{E-mail:\ }}
\end{aug}

\begin{abstract}

Ozone and particulate matter PM$_{2.5}$ are co-pollutants that have long been associated with increased public health risks.  Information on concentration levels
for both pollutants come from two sources: monitoring sites and output from complex numerical models that produce concentration surfaces over large spatial regions. In this paper,
we offer a fully-model based approach for fusing these two sources of information for the pair of co-pollutants which is computationally feasible over large spatial regions and long periods of time.  Due to the association between concentration levels of the two environmental contaminants, it is expected that information regarding one will help to improve prediction of the other.  Misalignment is an obvious issue since the monitoring networks for the two contaminants only partly intersect and because the collection rate for PM$_{2.5}$ is typically less frequent than that for ozone.

Extending previous work in Berrocal et al. (2010), we introduce a bivariate downscaler that provides a flexible class of bivariate space-time assimilation models.  We discuss computational issues for model fitting and analyze a dataset for ozone and PM$_{2.5}$ for the ozone season during year 2002.  We show a modest improvement in predictive performance, not surprising in a setting where we can anticipate only a small gain.

\end{abstract}



\end{frontmatter}

\section{Introduction}\label{sec:intro}

Ozone and particulate matter, PM$_{2.5}$, have long been associated with increased public health risks, e.g., of respiratory diseases
\cite{Schwartz1996,Dominici&2006,Braga&2001},
cardiovascular diseases \cite{Dominici&2006,Braga&2001},
and mortality and morbidity in general \cite{Dominici&2000,Smith&2000}. To set  air quality standards the EPA utilizes information from monitoring networks and from air quality computer models.

The two sources of information are valuable in different ways. The observations reported by monitoring networks, though sparsely collected and, sometimes, affected by missingness, provide direct measurements of the true value up to measurement error.
The output from air quality models estimates pollutant concentrations over large spatial domains at the grid cell level.  The estimates are viewed as averages over these cells and do not
contain any missing data, but are uncalibrated.
It is desirable to combine both sources of information but to do so, it is necessary to solve the inherent ``change of support'' problem of the data
\cite{Cressie1993, GotwayYoung2002,Banerjee&2004}, that is, the spatial misalignment between the observational data and the numerical model output.
Addressing the difference in spatial resolution between the two sources of data will also allow to evaluate and calibrate the numerical model output.

We are not interested in calibration of the
numerical model in the sense of \cite{KennedyOhagan2001}.  Rather our focus is on resolving the difference in spatial scale between the numerical model output
and the observational data, while bias-correcting the predictions generated by a numerical model.  In this regard, several methods have been proposed to assess numerical models and address the issue of ``incommensurability'' between grid model averages
and point measurements \cite{SwallFoley2009}.  Such effort is also common in the context of data assimilation within the atmospheric science literature, where the goal is to combine observational data on the current state of the atmosphere with a short-range forecast produced by a numerical weather prediction model to yield initial conditions for a numerical atmospheric model. The approaches tend to be algorithmic and ad hoc, only occasionally model-based, and do not necessarily address downscaling. Fuller discussion can be found in \cite{Daley1991} and \cite{Kalnay2003}.  In the context of air quality, \cite{Meiring&1998} propose to evaluate the hourly ozone predictions generated by a geophysical model
on the grid-cell scale by using the observations taken at monitoring sites, and predicting hourly level of ozone at the model grid cells by
averaging the predictions at $M$ regularly-spaced sites taken within each grid cell (see, as well, \cite{Fuentes&2003}).
A different strategy has been proposed by \cite{JunStein2004} who ignore the difference in spatial resolution between model output and observations, rather
suggesting to evaluate a numerical model by looking for differences between the model output and the observations in terms of variograms and correlograms.

\cite{FuentesRaftery2005} develop a ``fusion'' model, expressing the numerical model output
as the integral over a grid cell (scaled by the area of the cell) of a latent point-level process. Their main goal is to recover the true unobserved process, but, as a by-product, they obtain estimates of the bias of the numerical model output, thus
allowing evaluation and calibration of the numerical model. The work is an instance of Bayesian melding \cite{PooleRaftery2000},
and has gained considerable popularity \cite{SmithCowles2007,FoleyFuentes2008}. See also recent work in this spirit applied to sulfate aerosol \cite{SwallDavis2006} and ammonium \cite{DavisSwall2006}.
The Bayesian melding approach of \cite{FuentesRaftery2005}, though popular, suffers from several limitations, as pointed out by \cite{Liu&2007b} and \cite{Berrocal&2009}. Additionally,
it is only computationally feasible as a spatial model. A spatio-temporal extension, proposed by \cite{McMillan&2009}, specifies the latent process at the grid cell level, leading to comparison between the numerical model output
and the observational data at the grid cell scale.

Operating at the point level, \cite{Guillas&2008} propose to correct the model output by downscaling the predictions from grid cells to points and comparing them with the observations.
For each site, a two-step linear regression model that relates the observed hourly ozone level at a site with the numerical model output at the grid cell in
which the site lies is proposed. 
Building upon the work of \cite{Guillas&2008}, \cite{Liu&2008} suggest interpolating the intercept and the coefficient of the numerical model output, estimated at each site, via kriging,
thus developing a ``spatio-temporal'' model that will allow \emph{corrected} prediction of ozone level also at unmonitored sites. However, this ad-hoc approach does not provide estimates of the uncertainties associated with the predictions.
A formal spatio-temporal model, that builds upon the downscaling idea of \cite{Guillas&2008}, has been proposed as a univariate downscaler by \cite{Berrocal&2009} with a version implemented and described in \cite{Sahu&2009a}
and in \cite{GelfandSahu2009}.

The contribution of this paper is to extend the spatio-temporal downscaler from the univariate setting to a bivariate setting in order to exploit the correlation both in the observed concentration levels
of ozone and PM$_{2.5}$ and in the concentration levels of ozone and PM$_{2.5}$, provided by the numerical model.
Working in a bivariate setting will prove particularly advantageous for predicting particulate matter. The sampling frequency of the
PM$_{2.5}$ monitors - most of the monitors measure PM$_{2.5}$ concentration every 3 days - yields challenging interpolation of
the monitoring data to the entire spatial domain as well as difficult evaluation of the predictions generated by the numerical model.
Rather, in environmental health studies, researchers always use monitoring data to characterize exposure to fine particulate matter,
dealing with the PM$_{2.5}$ sampling frequency issue by aggregating the monitoring data over time
and/or over space. Similarly, most of the research effort on PM$_{2.5}$ has been devoted to developing spatial and spatio-temporal models to predict PM$_{2.5}$ based on monitoring data, meteorological covariates or observed concentrations for another pollutant \cite{Brown&1994,Le&1997,Kibria&2002,Smith&2003,SahuMardia2005,Sahu&2006}.
A fusion model to combine monitoring data for PM$_{2.5}$ with satellite aerosol optical depth (AOD) data has been proposed by \cite{PaciorekLiu2008}: however, the analysis
was conducted on monthly average concentration rather than daily and it revealed that AOD provides no improvement in predicting fine particulate matter.

In this paper, we present a general bivariate spatio-temporal downscaler developed for the numerical model outputs of an air quality model; we illustrate the methodology 
with regard to prediction of concentration of ozone and fine particulate matter. The model, which to our knowledge, is the first bivariate fusion/downscaler model, extends the univariate downscaler of \cite{Berrocal&2009}, and,
following \cite{Berrocal&2009}, regresses the bivariate vector of observed ozone and PM$_{2.5}$ concentration on the numerical model outputs for both
pollutants using spatially-varying coefficients \cite{SchmidtGelfand2003,Gelfand&2004}, modeled as a correlated six-dimensional Gaussian process.
The model specification is general and flexible, while offering feasible computation for a fully model-based fusion across space and time. We explore several different special cases, corresponding to different assumptions on the correlation structure of the two pollutants.
An important feature of our model is that it allows us to handle not only the spatial misalignment between monitoring data and model output, but also it allows us to accommodate the spatial and temporal misalignment between the ozone and PM$_{2.5}$ monitoring data.
Finally, exploiting the correlation, not only between the model output and the corresponding monitoring
data for a pollutant, but also between pollutants, we can jointly predict ozone and PM$_{2.5}$ at any site in the spatial domain and provide a measure of the uncertainties associated with such predictions.
Additionally, at sites where one pollutant is measured but the other is not, we can predict the level of the latter.

The paper is organized as follows. In Section 2, we present the monitoring data and the numerical model output, in Section 3 we describe the bivariate downscaler model,
by first introducing the downscaler in the univariate setting, and then extending the model to the bivariate case, first in a purely spatial setting and then in a spatio-temporal setting.
Details on how to handle the spatio-temporal misalignment in the monitoring data are also discussed in Section 3. Section 4 presents details on the model fitting, while Section 5
displays results of our analysis. Finally, we conclude in Section 6
with a discussion and evaluation of our method and with indications for future work.

\bigskip

\section{Data}\label{sec:data}

Particulate matter (PM$_{2.5}$) and ozone are two of the ``criteria pollutants" that the Environmental Protection Agency (EPA) is required to monitor by the Clean Air Act.
The first is a mixture of solid and liquid particles, emitted in the atmosphere either directly from a source
or as result of complicated reactions of chemicals, while the second is a gas made of three atoms of oxygen that is created
by a chemical reaction between oxides of nitrogen and volatile organic compounds in the presence of sunlight.

The EPA tracks both criteria pollutants using both measurements taken at monitoring sites and estimates of air pollutants concentration produced by the
Models-3/Community Mesoscale Air Quality (CMAQ) model \cite{ByunSchere2006}(\texttt{epa.gov/asmdnerl/CMAQ}).
The latter is a deterministic numerical model that predicts concentrations
for various pollutants by integrating three major components: an atmospheric component accounting for the atmosphere and its states and motions,
an emission component accounting for the emissions injected in the atmosphere, and
a chemical component accounting for the reactions between the different gas present in the atmosphere.
By simulating various chemical and physical processes, such as horizontal and vertical advection, emission injection, deposition, plume chemistry effects,
etc., CMAQ produces estimates of pollutants concentration at pre-determined spatial scales, e.g., 36 km, 12 km, and, most recently, at 4 km grid cells.

\begin{figure}[!t]
\centering
\includegraphics[scale=0.35,angle=0]{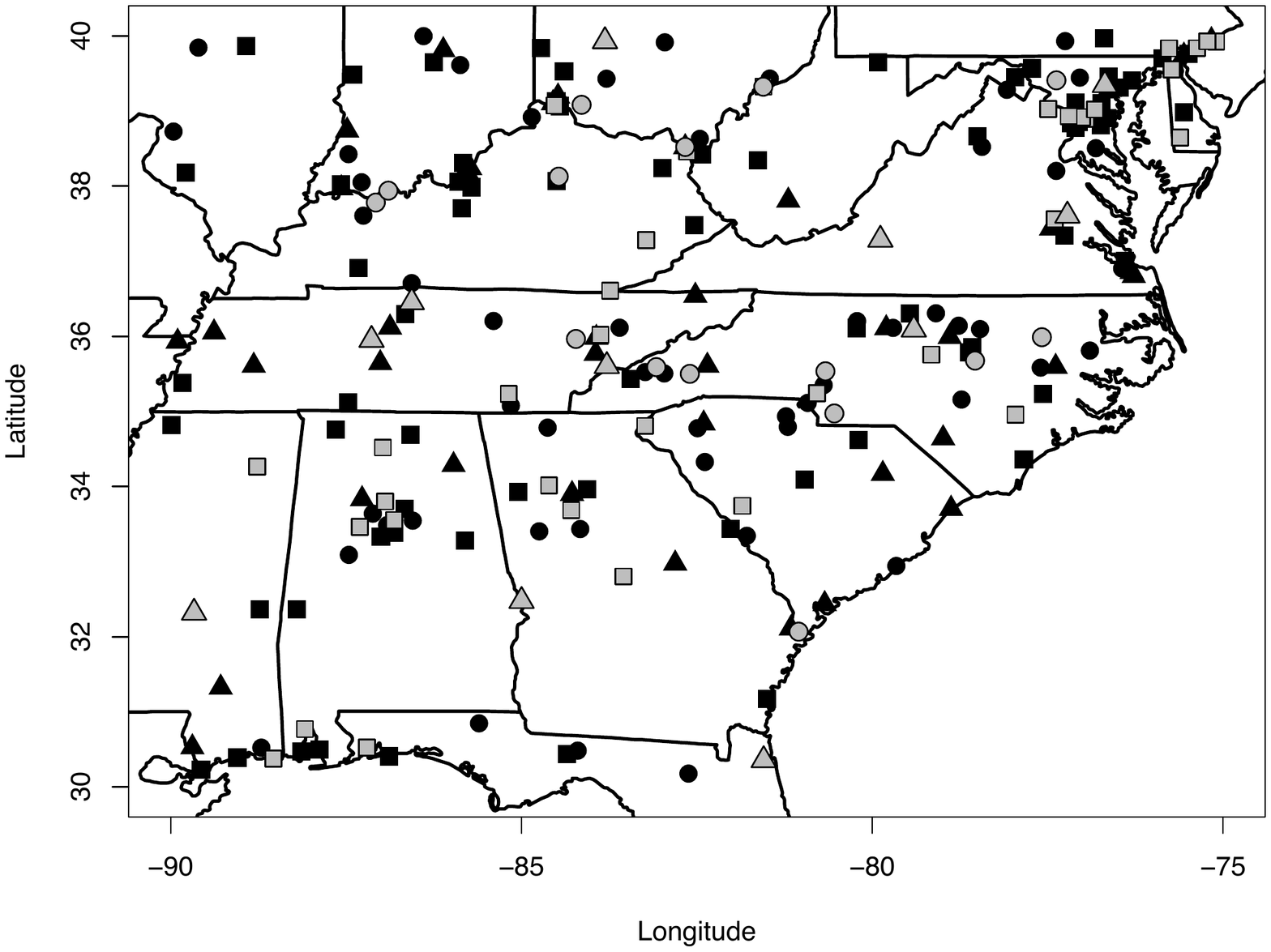}
\vspace{0.2cm}
\caption{Sites reporting concentration of ozone and PM$_{2.5}$ used in our case study. Sites measuring only ozone are represented with dots, sites reporting only
PM$_{2.5}$ concentrations are represented with triangles, while sites measuring concentration for both pollutants are represented with squares.
Black symbols are used to display sites used to fit the model, while grey symbols indicate validation sites.
\label{fig:studyarea}}
\end{figure}

We consider daily estimates and measurements of ozone and PM$_{2.5}$ concentration for the region in the South Eastern part of the
United States shown in Figure~\ref{fig:studyarea} during the period June 1 - September 30, 2002, the summer season when solar radiation and temperature are high.  In turn this encourages not only high concentrations of ozone but also of particulate SO$_{4}$, the dominant component of PM$_{2.5}$ in the eastern U.S as well as particulate ammonium nitrate.  Relatively strong association between ozone and PM$_{2.5}$ concentrations is expected and, in fact, observed.
Following the air quality standards (NAAQS; http://www.epa.gov/air/criteria.html) set by EPA for ozone and PM$_{2.5}$, daily ozone concentration is measured as the daily maximum
8-hour average concentration, while daily PM$_{2.5}$ concentration is given by the 24-hour average concentration of particulate matter.

The monitoring data used in our case study comes from 226 sites sparsely located in the region, and it has been obtained from the EPA Air Quality System (AQS) repository database.
Of these 226 sites, not all measure both pollutants: 71 report only ozone measurements, 50 report only PM$_{2.5}$, and 105 measure both, pinpointing the spatial misalignment
in the monitoring data.
In order to validate the model with out-of-sample predictions, we
randomly split the sites into two sets: a training set used to fit the model, and a validation set used to assess the performance of the model.  In the training set, 52 sites measured only ozone concentration, 39 reported only PM$_{2.5}$ concentration, and 70 reported both. In total, the dataset used to fit the model comprised 14,630 \emph{daily} measurements
of ozone concentration and 4,790 measurements of PM$_{2.5}$ concentration. The smaller number of observations for particulate matter is due to the
sampling scheme for PM$_{2.5}$: of the 109 PM$_{2.5}$ monitoring sites, only 11 (10.1$\%$) measure concentration of particulate matter every day,
while 90 (82.5$\%$) report measurements every three days, and 8 (7.3$\%$) measure PM$_{2.5}$ every six days.

The difference in sampling frequency between ozone and PM$_{2.5}$ suggests that, in addition to spatial misalignment,
there is also temporal misalignment in the monitoring data: even sites that measure both pollutants do not sample the two pollutants with the same temporal frequency.
However, when concentration measurements for both pollutant are available, they display considerable correlation, equal to 0.69.
Spatial and temporal misalignment is also present among the validation sites: 54 measured only ozone, 46 only PM$_{2.5}$ sites, while
35 reported concentration for both pollutants. In total,
the validation dataset contained 6,530 daily observations of ozone concentration and 2,559 daily observations of concentration of particulate matter.
The location of the sites used to fit the and validate the model is shown in Figure~\ref{fig:studyarea}.

\begin{figure}[!ht]
\centering
\subfigure[]{\includegraphics[scale=0.35,angle=0]{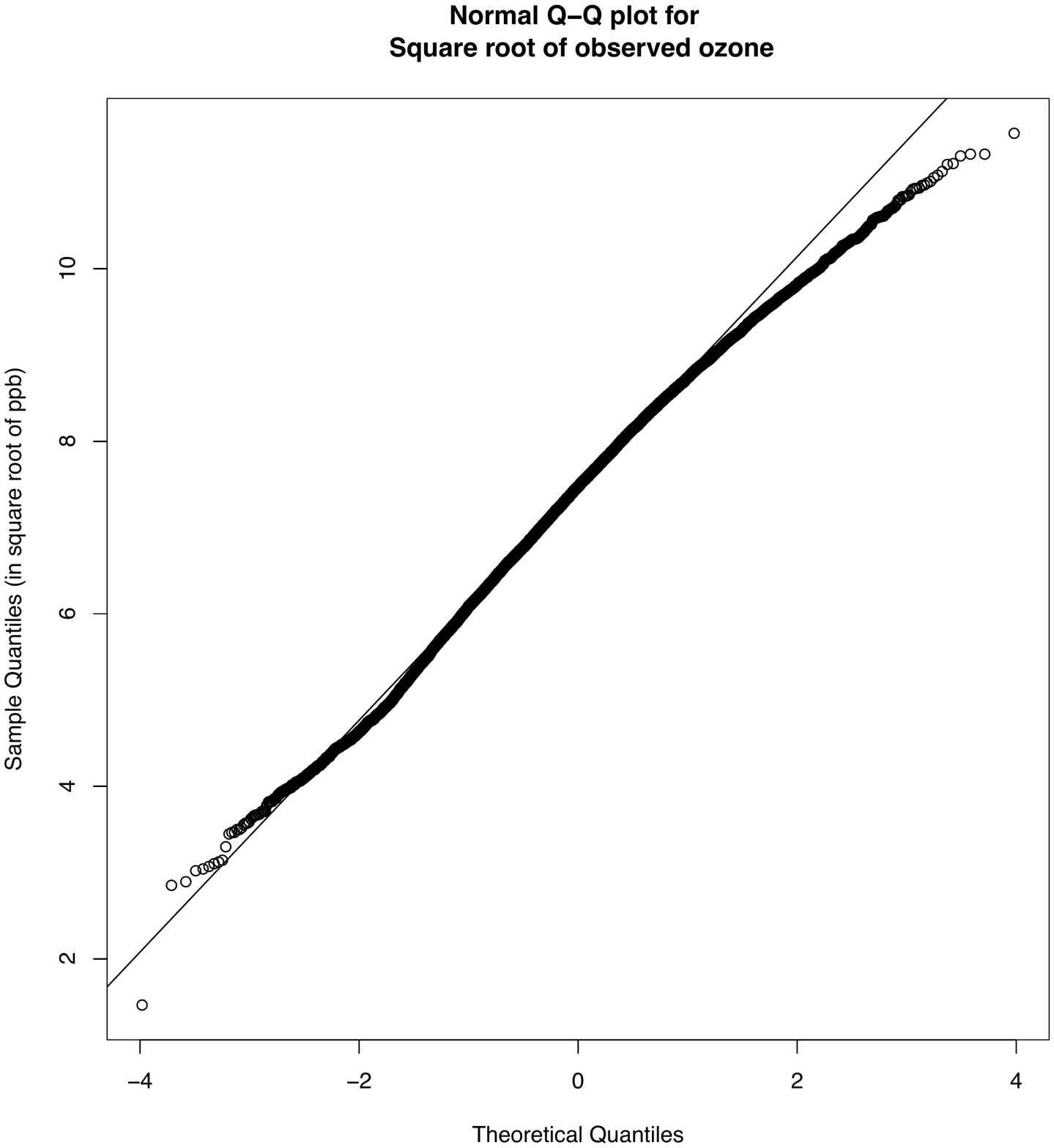}}
\qquad
\subfigure[]{\includegraphics[scale=0.35,angle=0]{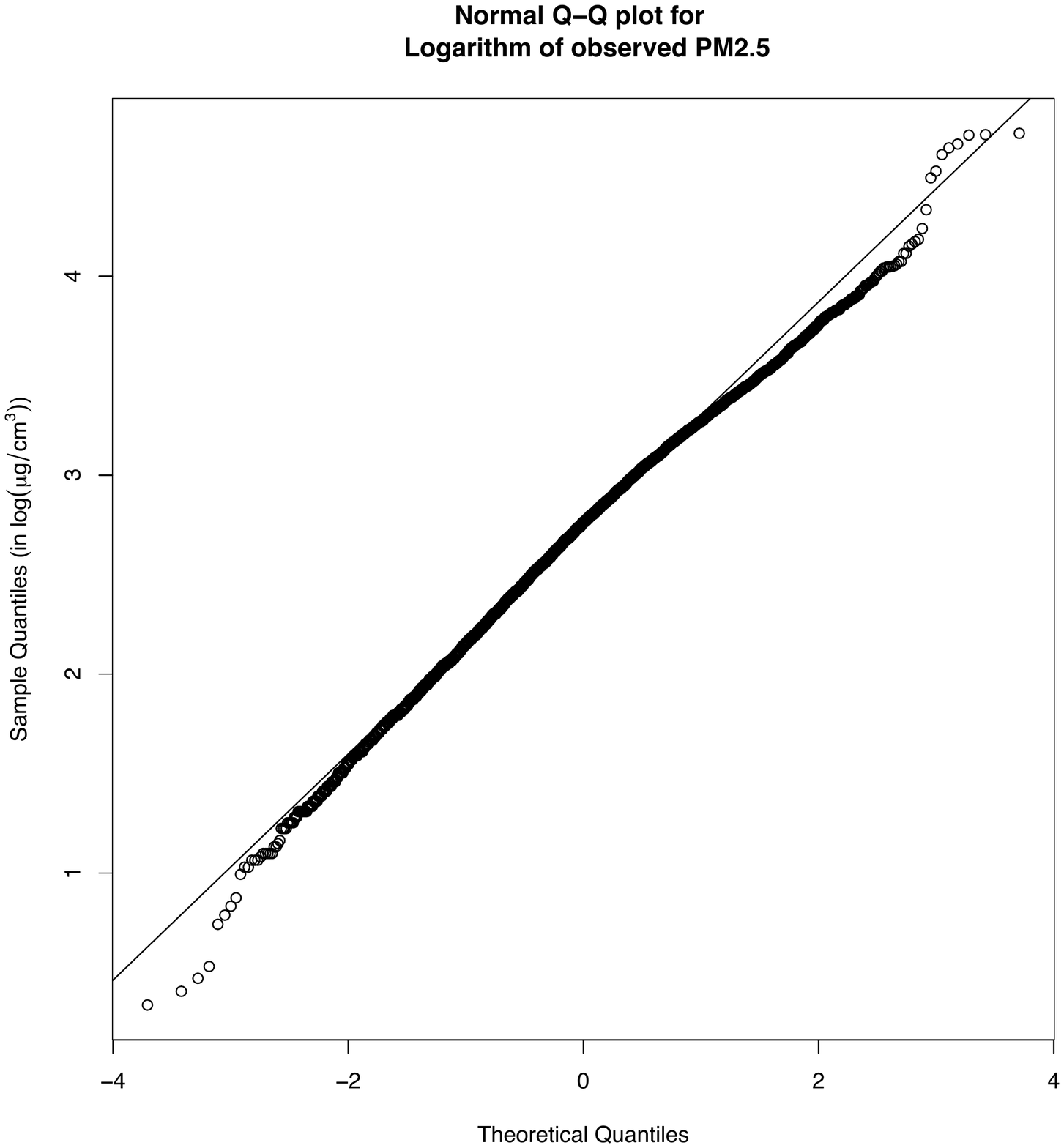}}
\vspace{0.2cm}
\caption{Normal Q-Q plots of: (a) square root of observed ozone; (b) log of observed PM$_{2.5}$.
\label{fig:histtest}}
\end{figure}

An exploratory analysis of the observed daily concentration data for ozone and PM$_{2.5}$ revealed the need for a transformation. We modeled the
daily ozone concentration data on the square root scale to stabilize the variance, while we log-transformed the PM$_{2.5}$
concentration data to achieve normality. Both transformations have been previously used; see, e.g., \cite{Sahu&2007}, \cite{Sahu&2006}, \cite{Carroll&1997},
\cite{HaslettRaftery1989}.
A normal Q-Q plot of the square root of the observed daily concentration of ozone at sites used to fit the model is shown in Figure~\ref{fig:histtest} (a), while Figure~\ref{fig:histtest} (b) presents
a normal Q-Q plot of the logarithm of daily average concentration of PM$_{2.5}$ at the 109 monitoring sites in the test set measuring particulate matter.  Both plots seems to suggest a slight deviation from normality, however the deviation is minimal and we are comfortable using a normal model, in agreement with the literature.
Neverthless, in Section \ref{sec:summary} we address distributional issues more at lengths and suggest an extension of our downscaling approach for non-Gaussian
variables.
The overall mean and standard deviation for the square root of ozone concentration and for the log of PM$_{2.5}$ concentration at the test sites
were, respectively, 7.41 and 1.31 $\sqrt{\mbox{ppb}}$ (ppb:parts per billions), and 2.72 and 0.56 $\log(\mu \mbox{g/m}^{3})$.

\begin{figure}[!ht]
\centering
\subfigure[]{\includegraphics[scale=0.35,angle=0]{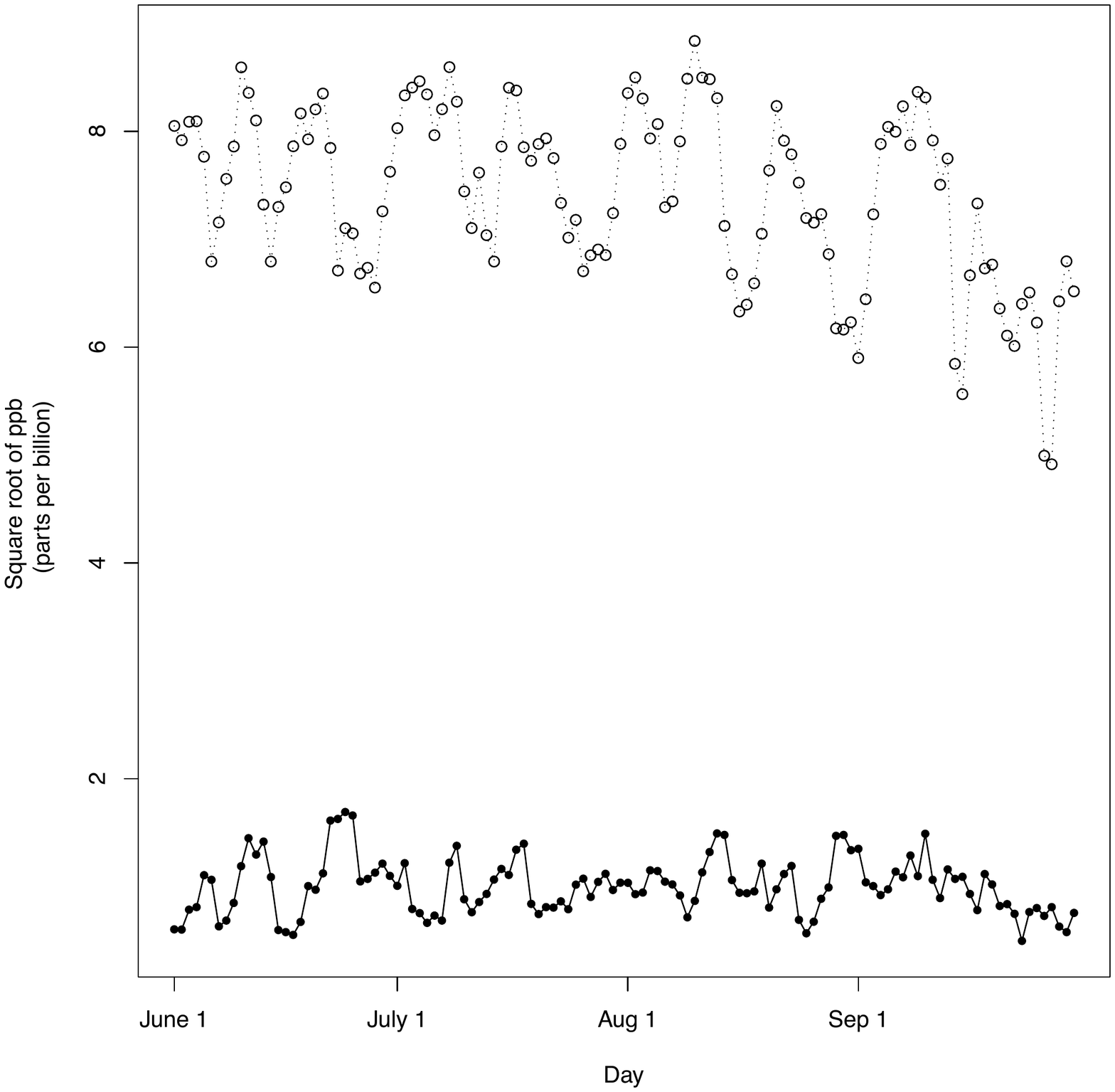}}
\qquad
\subfigure[]{\includegraphics[scale=0.35,angle=0]{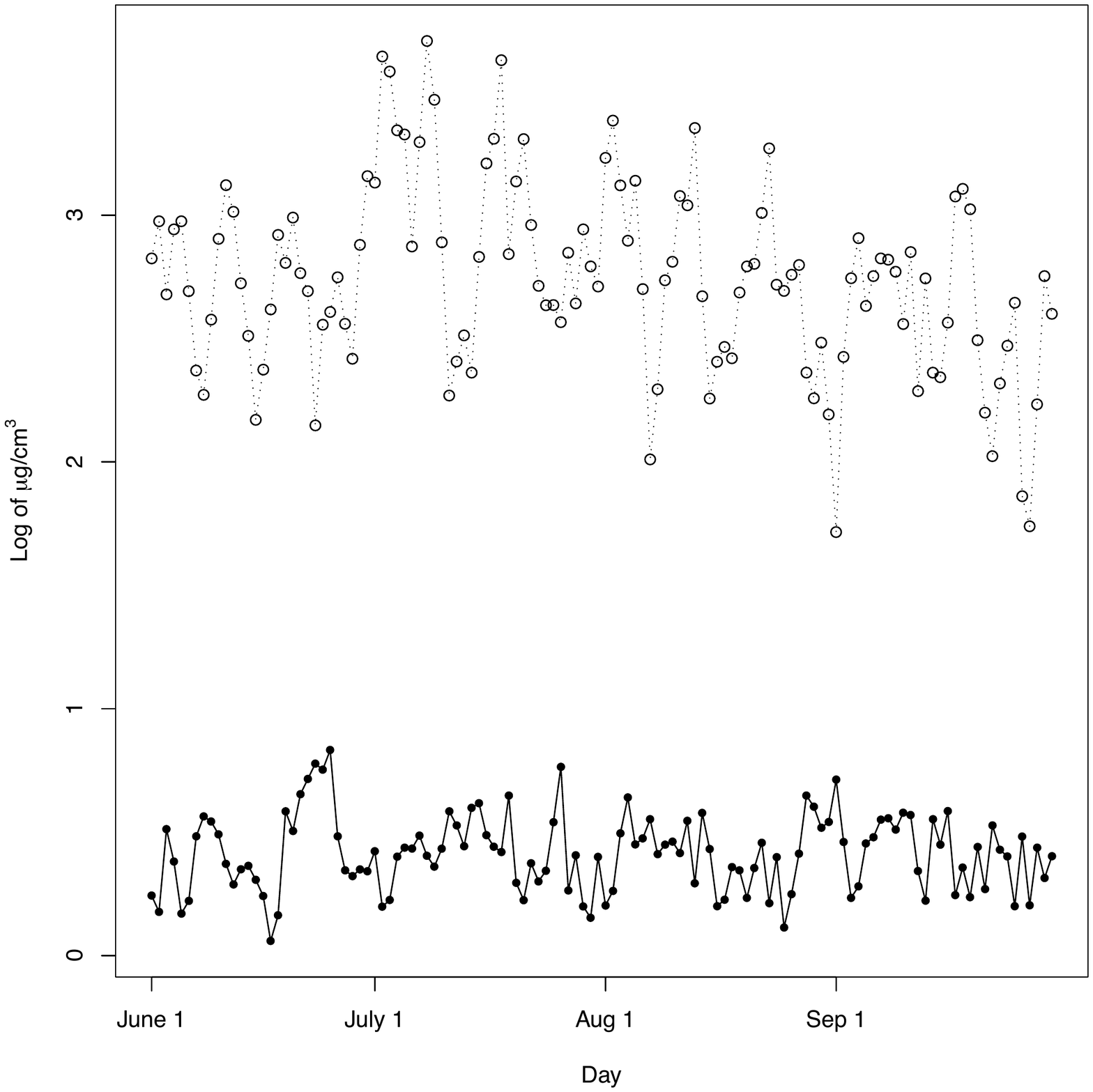}}
\vspace{0.2cm}
\caption{Time series of: (a) daily mean (open circles) and daily standard deviation (black dots) of square root of observed ozone; (b) daily
mean (open circles) and daily standard deviation (black dots) of log of observed PM$_{2.5}$.
\label{fig:timeseries}}
\end{figure}

Estimates of the daily average and standard deviation for ozone and PM$_{2.5}$ are presented in Figure~\ref{fig:timeseries}. Both panels in Figure~\ref{fig:timeseries}
indicate daily variability in the concentration level for the pollutants, with Figure~\ref{fig:timeseries}(a) revealing the seasonality of ozone during the study period (June 1 - September 20, 2002). In terms of variability, both ozone and PM$_{2.5}$
present the largest standard deviation on June 25, 2002.
For this day, plots of the observed ozone and PM$_{2.5}$ concentration are shown, on the original scales, respectively, in Figure~\ref{fig:ozonejun25}(a) and Figure~\ref{fig:pmjun25}(a).  (These plots are developed in conjunction with the data analysis of Section 5.)

The numerical model output data consists of estimates of ozone and PM$_{2.5}$ concentration level generated by the CMAQ numerical
model run at 12 km grid cell resolution. An example of the type of concentration surfaces yielded by CMAQ can be observed, respectively, in Figure~\ref{fig:ozonejun25}(b) and Figure~\ref{fig:pmjun25}(b).
CMAQ produces estimates of a pollutant concentration in terms of average over a grid cell (in this case, over 10,504 12-km grid cells), while observations are taken at points. Therefore, there is a spatial misalignment between observational and numerical model data.
We solve this difference in spatial resolution between the monitoring and the numerical model data by associating to each site $\mathbf{s}$ in the spatial domain $\mathcal{S}$, the CMAQ grid cell $B$
in which $\mathbf{s}$ lies. This allows a comparison between the model output and the monitoring data.

To reveal the need for calibration in the numerical model output, we have produced pairwise scatterplots of the square root of the observed ozone concentration versus the square root of the CMAQ predicted concentration level of ozone during the period June 1 - September 30, 2002, of the square root of the observed
ozone concentration versus the log of the CMAQ predicted PM$_{2.5}$ concentration, and analogous plots for the log of the observed
concentration of particulate matter. We see that the predictions by the numerical model are biased and need to be calibrated;
however, they do contain useful information to improve prediction for both pollutants.
Additionally, the positive and substantial correlation between the CMAQ model output for PM$_{2.5}$ and the observed ozone concentration (r=0.62),
and, similarly, between the CMAQ model output for ozone and the observed PM$_{2.5}$ concentration (r=0.69), indicate that the CMAQ output for PM$_{2.5}$
might be useful to predict ozone and conversely.

\begin{figure}[!hp]
\centering
\begin{tabular}{cc}
\includegraphics[scale=0.3,angle=0]{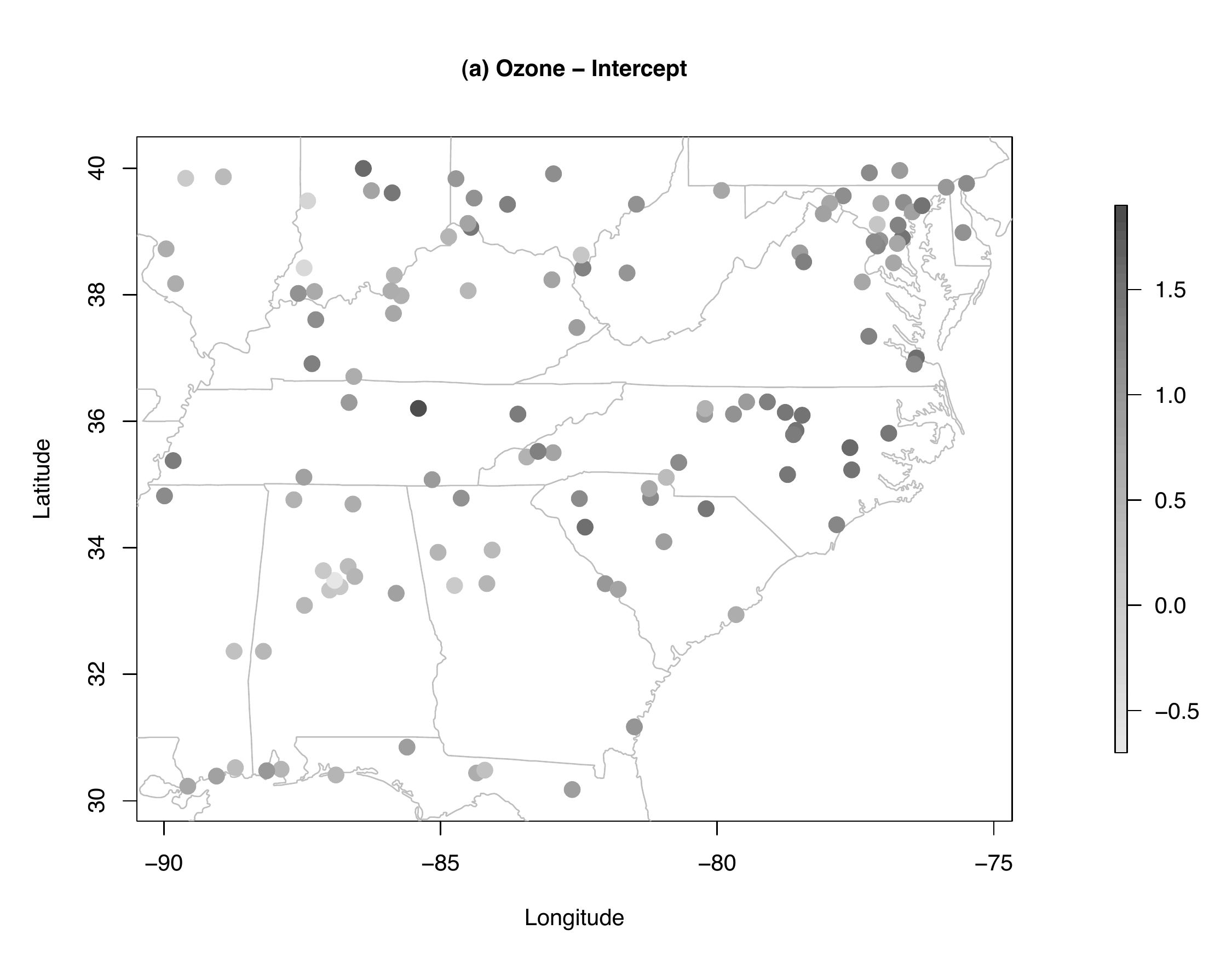}
&
\includegraphics[scale=0.3,angle=0]{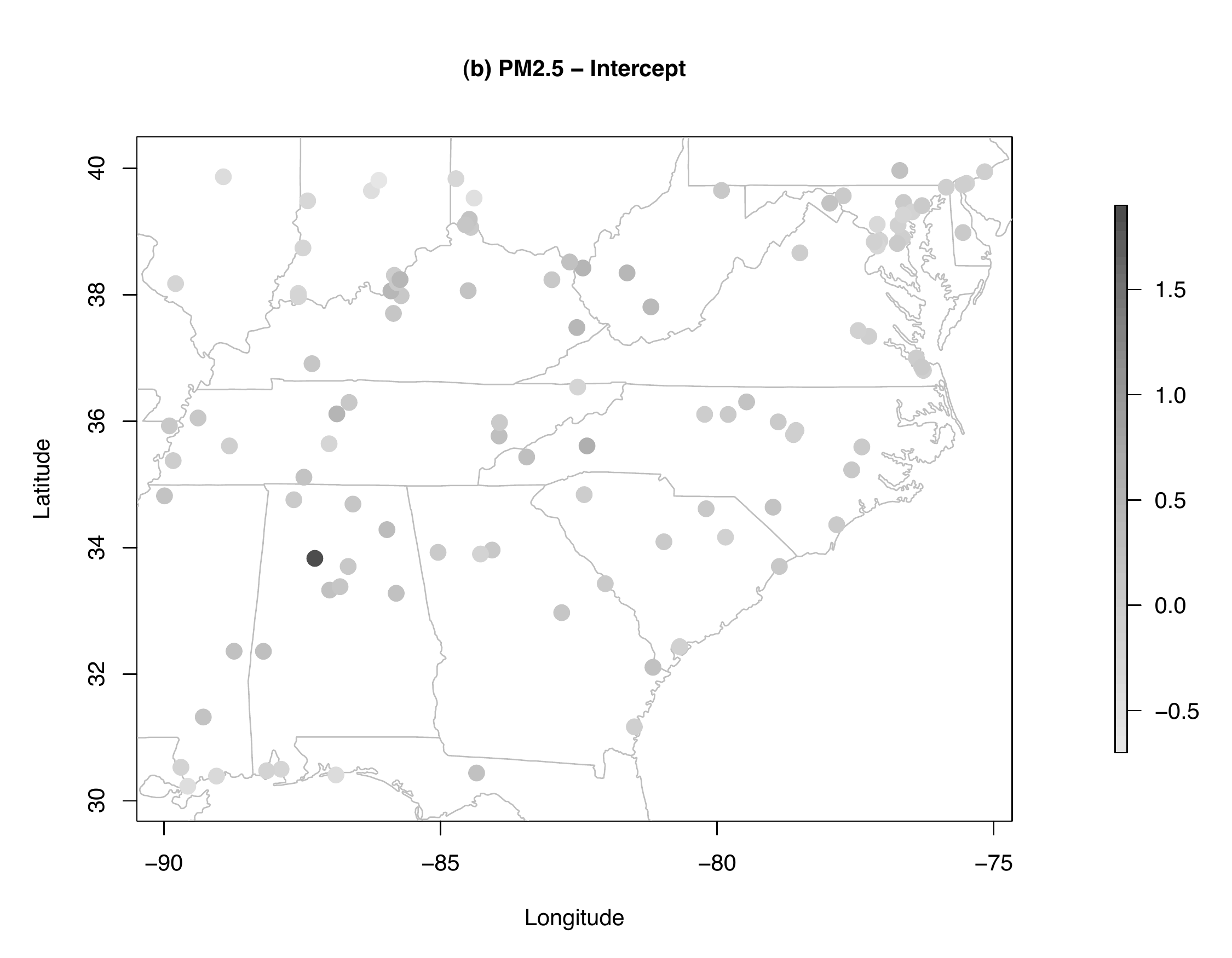}
\\
\includegraphics[scale=0.3,angle=0]{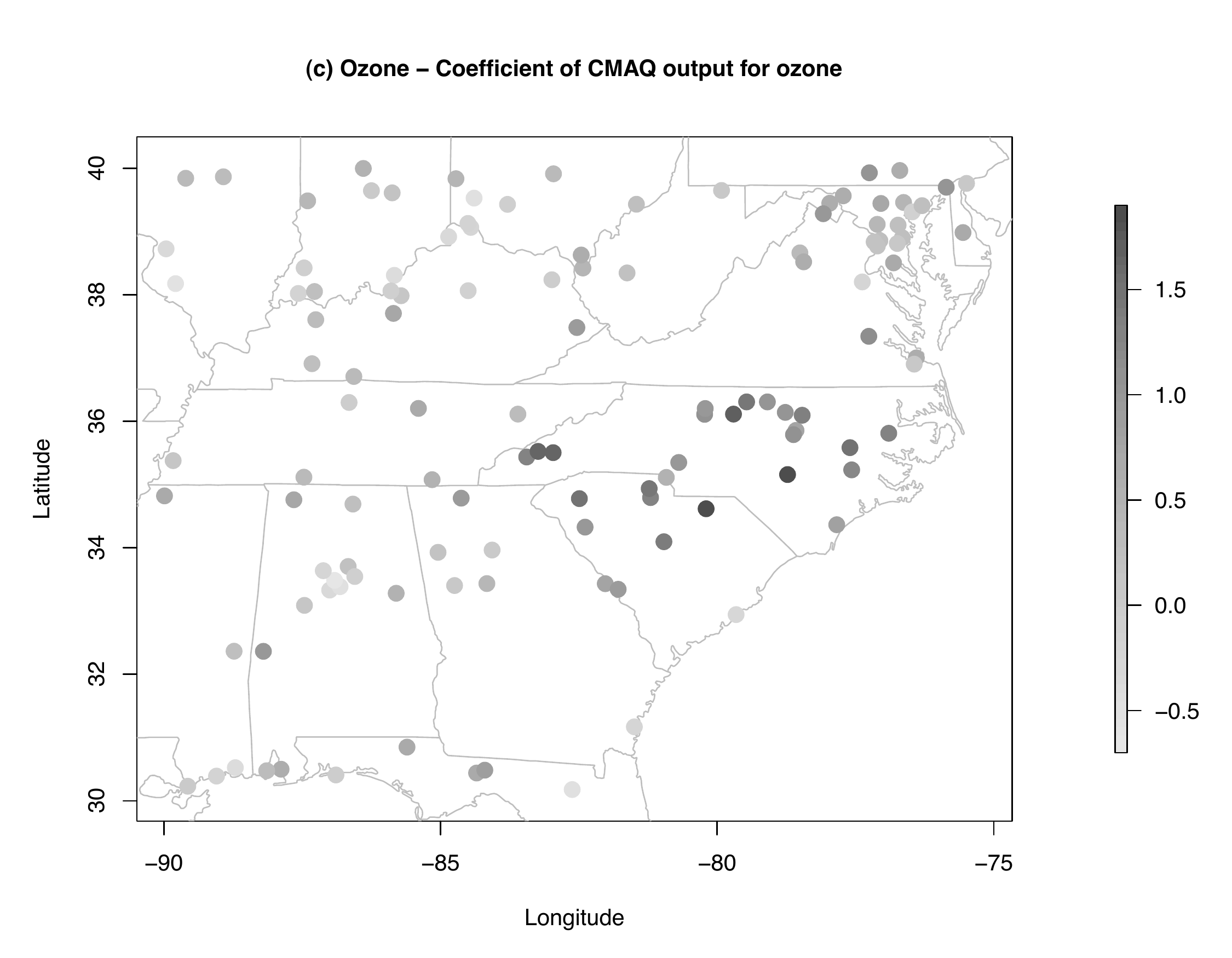}
&
\includegraphics[scale=0.3,angle=0]{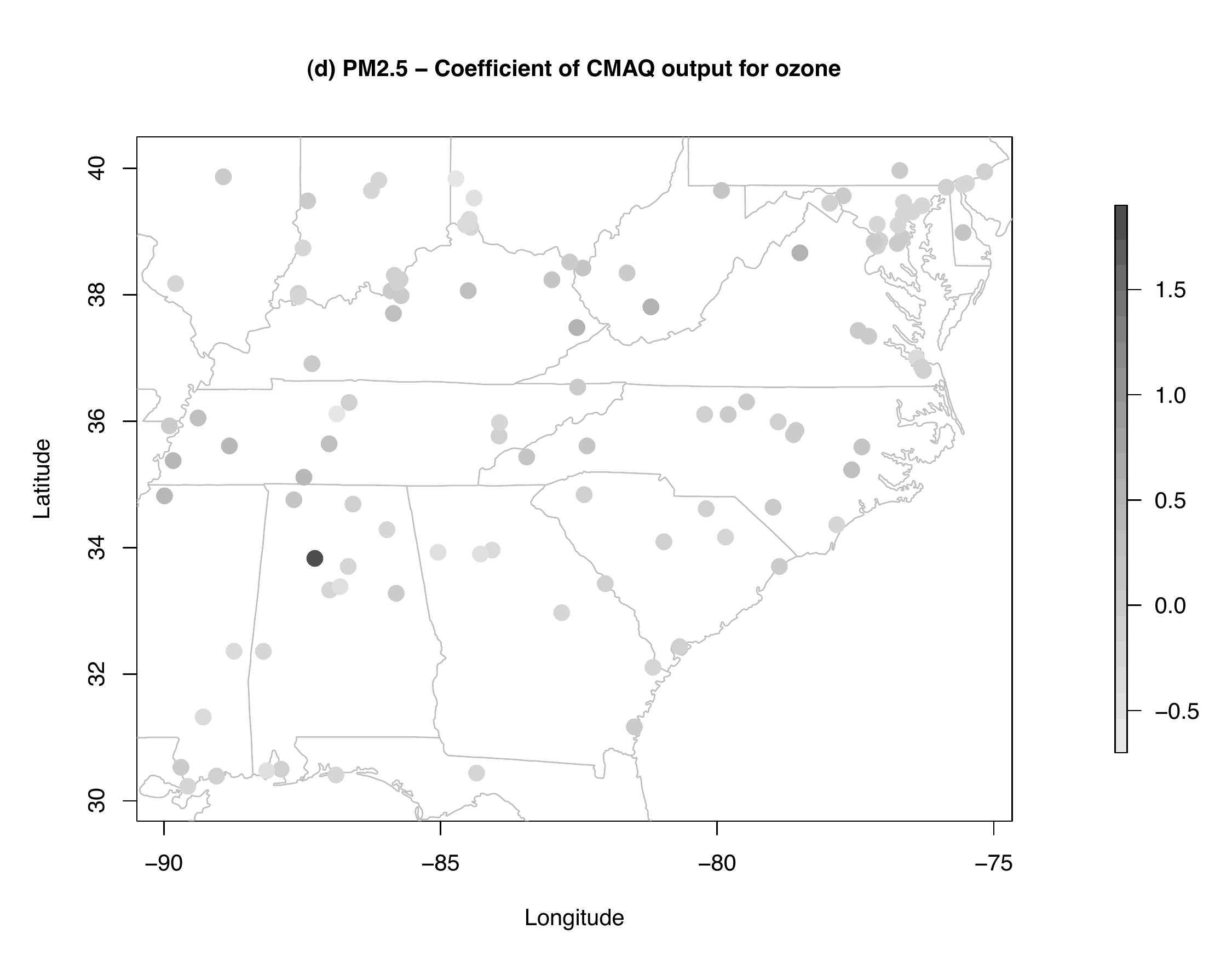}
\\
\includegraphics[scale=0.3,angle=0]{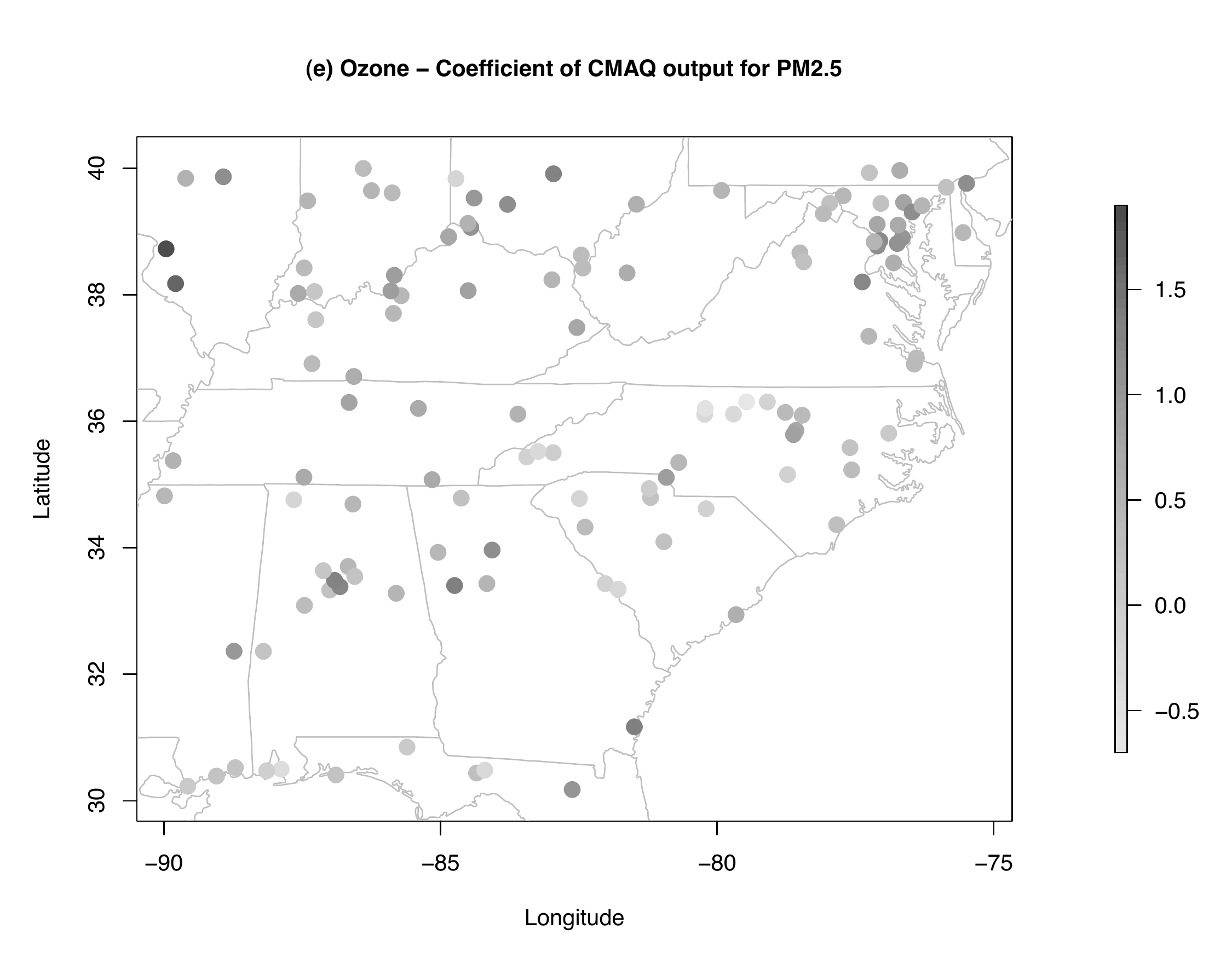}
&
\includegraphics[scale=0.3,angle=0]{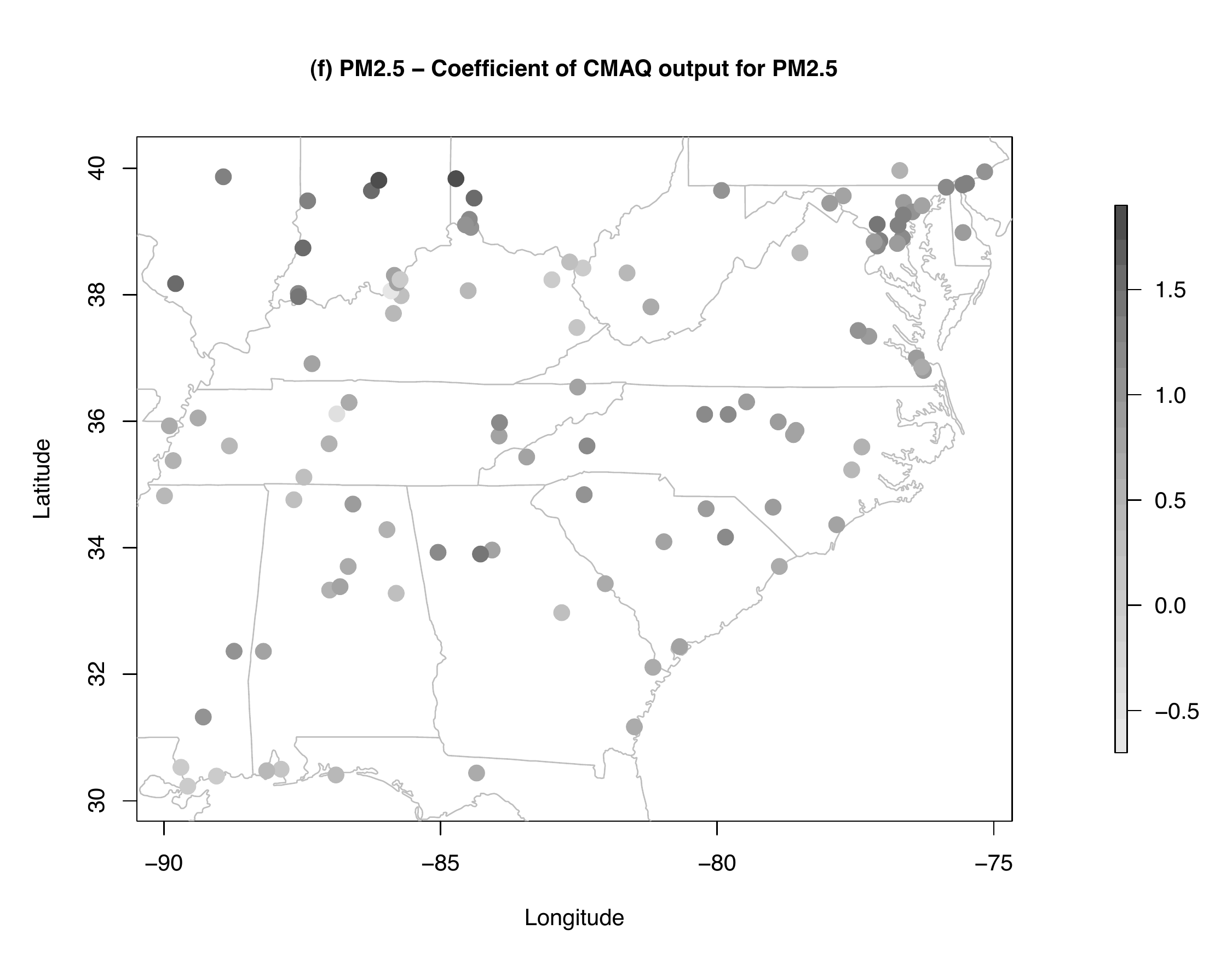}
\\
\end{tabular}
\caption{Spatial maps of the estimates of the coefficients of the linear regressions of the square root of
observed concentration of ozone (panels (a), (c) and (e)) and of the logarithm of the observed concentration of PM$_{2.5}$
(panels (b), (d) and (f)) on the square root of the CMAQ predicted concentration of ozone and on the log
of the CMAQ predicted concentration of PM$_{2.5}$: (a)-(b) intercept term; (c)-(d) coefficient
of the CMAQ model output for ozone; (e)-(f) coefficient of the CMAQ model output for PM$_{2.5}$.
In all panels, the linear regression has been carried out between the normalized response and the normalized covariates.
\label{fig:est_coeff_map}}
\end{figure}

The type of calibration implied by the pairwise scatterplots mentioned above is constant across the entire spatial domain $\mathcal{S}$
in consideration. Empirical evidence suggests instead that the error and the bias of the numerical model output might not be constant in space, rather
they vary from site to site. Figure~\ref{fig:est_coeff_map} presents spatial maps of the estimates of the intercept and of the regression coefficients of 
CMAQ ozone and CMAQ PM$_{2.5}$
at each of the sites used to fit our bivariate downscaler model. These estimates have been obtained by regressing at each site $\mathbf{s}$, respectively, the normalized
log of the observed PM$_{2.5}$ concentration at $\mathbf{s}$ across time
on the corresponding normalized 
CMAQ model output for ozone and PM$_{2.5}$, both taken on the appropriate scale (square root for ozone, log for PM$_{2.5}$).
As the plots indicate, there is spatial variability in the estimates of these coefficients with differential variability across the estimates.
In particular, we see less spatial variability in the estimates obtained for the observed log PM$_{2.5}$ concentration. Also, there is a difference
between the two pollutants in the significance of the estimates of the coefficients: while for ozone, the estimates of the coefficient of CMAQ ozone are all significant, and
about 60$\%$ of the estimates of the intercept are significant, in the regression for PM$_{2.5}$, only 21$\%$ of the intercepts are significantly different from zero,
and about 88$\%$ of the estimates of the coefficient of CMAQ PM$_{2.5}$ are significantly different from zero.

\section{Modeling}\label{sec:model}

Here we present the model underlying our downscaling technique.
We first briefly review the univariate downscaler, presented in \cite{Berrocal&2009}, then we extend it to the bivariate setting. In doing so, we
first present the general bivariate downscaler in a purely spatial setting, then we examine extension to accommodate
data collected over time.

\subsection{Univariate downscaler}\label{subsec:univariate}

Let $Y(\mathbf{s})$ be the observed data at $\mathbf{s}$ and let $x(B)$ be the numerical model output at grid cell $B$. Then, $x(B)$ is interpreted as the average level of the
variable under consideration over $B$. The spatial downscaler addresses the difference in spatial resolution
between the observational data and the numerical model output, i.e. the ``change of support'' problem  \cite{Cressie1993,Banerjee&2004},
by associating to each point $\mathbf{s}$ the grid cell $B$ in which $\mathbf{s}$ lies. Then, it relates
the observational data and the numerical model output as follows:
\begin{equation}
Y(\mathbf{s}) = \tilde{\beta}_{0}(\mathbf{s}) + \tilde{\beta}_{1}(\mathbf{s}) x(B) + \epsilon(\mathbf{s}) \qquad \epsilon(\mathbf{s}) \stackrel{iid}{\sim} N(0,\tau^{2})
\label{eq:unidown_eq}
\end{equation}
where
\begin{eqnarray}
\tilde{\beta}_{0}(\mathbf{s}) & = & \beta_{0} + \beta_{0}(\mathbf{s}) \nonumber \\
\tilde{\beta}_{1}(\mathbf{s}) & = & \beta_{1} + \beta_{1}(\mathbf{s})
\label{eq:unidown_coeff}
\end{eqnarray}
In (\ref{eq:unidown_coeff}), $\beta_{0}$ and $\beta_{1}$ denote, respectively, the overall additive and multiplicative bias of the numerical model, and
$\beta_{0}(\mathbf{s})$ and $\beta_{1}(\mathbf{s})$ are local adjustments to the overall bias terms.
Finally, $\epsilon(\mathbf{s})$ is a white noise process with nugget variance $\tau^{2}$. Evidently, a constant-mean downscaler is
a special case of (\ref{eq:unidown_coeff}) obtained when the local adjustments, $\beta_{0}(\mathbf{s})$ and $\beta_{1}(\mathbf{s})$, are set to zero.

Anticipating association between intercept and slope, at the second hierarchical level of the model, we set the two spatially-varying coefficients $\beta_{0}(\mathbf{s})$ and $\beta_{1}(\mathbf{s})$ to be
a bivariate correlated mean-zero Gaussian processes. Using the method of coregionalization
\cite{Wackernagel1998,SchmidtGelfand2003,Gelfand&2004},
we assume that there exist two latent mean-zero, unit-variance independent Gaussian processes $w_{0}(\mathbf{s})$ and $w_{1}(\mathbf{s})$ such that
$\mbox{Cov}(w_{j}(\mathbf{s}),w_{j}(\mathbf{s}^{\prime}))=\exp (-\phi_{j} \| \mathbf{s} - \mathbf{s}^{\prime} \|)$, where, for $j=0,1$,
$\phi_{j}$ is the spatial decay parameter of the Gaussian process $w_{j}(\mathbf{s})$, $\| \mathbf{s} - \mathbf{s}^{\prime} \|$ is the great circle distance between $\mathbf{s}$
and $\mathbf{s}^{\prime}$ \footnote{When considering large spatial domains, it is preferable to use three-dimensional Euclidean distance as an argument for the exponential covariance function rather
than great-circle distance, which might yield a covariance matrix which is not positive definite. For the domain we study, the two metrics almost coincide and yield very similar results.}, and
\begin{equation}
\left(
\begin{array}{c}
\beta_{0}(\mathbf{s})  \\
\beta_{1}(\mathbf{s})
\end{array}
\right)
= \mathbf{A}
\left(
\begin{array}{c}
w_{0}(\mathbf{s}) \\
w_{1}(\mathbf{s})
\end{array}
\right)
\label{eq:unidown_coreg}
\end{equation}

The coregionalization matrix $\mathbf{A}$ in (\ref{eq:unidown_coreg}) determines the correlation between the two spatially-varying coefficients
$\beta_{0}(\mathbf{s})$ and $\beta_{1}(\mathbf{s})$ and can be assumed, without loss of generality, to be lower-triangular. Note that $\mathbf{T}= \mathbf{A}\mathbf{A}^{\prime}$
is the constant local covariance matrix.  The specification of the
model is then completed with the priors for: the overall bias terms, $\beta_{0}$ and $\beta_{1}$,
the nugget variance $\tau^{2}$, the three non-null elements of the coregionalization matrix $\mathbf{A}$, and the decay parameters $\phi_{0}$
and $\phi_{1}$.  Other choices for obtaining dependent Gaussian processes could be entertained, including moving averages \cite{VerHoefBarry1998}, convolution of covariance functions \cite{MajumdarGelfand2007}, and particular parametric choices. In the last case, some possibilities include
 the cross-covariance functions introduced by \cite{ApanasovichGenton2009} and \cite{Gneiting&2009}.

\subsection{The bivariate downscaler: static version}\label{subsec:bivariate}

Turning to the bivariate case, we describe the model in terms of square root ozone and log PM$_{2.5}$ concentrations, recognizing that the model could be applied to any suitable pair of point referenced variables. Let $Y_{1}(\mathbf{s})$ and
$Y_{2}(\mathbf{s})$ denote, respectively, the square root and the logarithm of the observed ozone and PM$_{2.5}$ concentration at $\mathbf{s}$, and let
$x_{1}(B)$ and $x_{2}(B)$ indicate, respectively, the square root and the logarithm of the numerical model output for ozone and PM$_{2.5}$ at grid cell $B$.
Then, $\mathbf{Y}(\mathbf{s})=(Y_{1}(\mathbf{s}),Y_{2}(\mathbf{s}))$ is the bivariate random vector with the observed concentrations for the two pollutants at $\mathbf{s}$, while
$\mathbf{x}(B)=(x_{1}(B),x_{2}(B))$ is the bivariate vector with the two numerical model outputs.
Building upon (\ref{eq:unidown_eq}), our bivariate downscaler model relates the monitoring data at $\mathbf{s}$ and the numerical model outputs at grid cell $B$, with $\mathbf{s}$ lying in $B$, as follows:
\begin{eqnarray}
Y_{1}(\mathbf{s}) & = & \tilde{\beta}_{10}(\mathbf{s}) + \tilde{\beta}_{11}(\mathbf{s}) x_{1}(B) + \tilde{\beta}_{12}(\mathbf{s}) x_{2}(B) + \epsilon_{1}(\mathbf{s}) \nonumber \\
Y_{2}(\mathbf{s}) & = & \tilde{\beta}_{20}(\mathbf{s}) + \tilde{\beta}_{21}(\mathbf{s}) x_{1}(B) + \tilde{\beta}_{22}(\mathbf{s}) x_{2}(B) + \epsilon_{2}(\mathbf{s})
\label{eq:bidown_eq}
\end{eqnarray}
where $\epsilon_{1}(\mathbf{s})$ and $\epsilon_{2}(\mathbf{s})$ are two independent white noise processes with nugget variances, respectively, $\tau^{2}_{1}$ and $\tau^{2}_{2}$.

As in the univariate case, we adopt a random effect notation, decomposing each of the $\beta_{ij}(\mathbf{s})$, $i=1,2$, $j=0,1,2$, into the sum of an overall term and a local adjustment.  That is,
\begin{equation}
\tilde{\beta}_{ij}(\mathbf{s}) = \beta_{ij} + \beta_{ij}(\mathbf{s}).
\label{eq:bidown_decomp}
\end{equation}

The six-dimensional process $(\beta_{ij}(\mathbf{s}))_{i=1,2; j=0,1,2}$ is in turn modeled as a six-dimensional mean-zero correlated Gaussian process, again using the method of coregionalization. Therefore,
we express each of the spatially-varying coefficients $\beta_{ij}(\mathbf{s})$ as a linear combination of mean-zero unit-variance latent independent Gaussian processes, each equipped with an exponential covariance structure. More specifically, assuming, without loss of generality, that the coregionalization matrix $\mathbf{A}$ is lower triangular, we have that:
\begin{equation}
\left(
\begin{array}{c}
\beta_{10}(\mathbf{s})  \\
\beta_{11}(\mathbf{s}) \\
\beta_{12}(\mathbf{s})  \\
\beta_{20}(\mathbf{s}) \\
\beta_{21}(\mathbf{s})  \\
\beta_{22}(\mathbf{s}) \\
\end{array}
\right)
= \mathbf{A}
\left(
\begin{array}{c}
w_{1}(\mathbf{s}) \\
w_{2}(\mathbf{s}) \\
w_{3}(\mathbf{s}) \\
w_{4}(\mathbf{s}) \\
w_{5}(\mathbf{s}) \\
w_{6}(\mathbf{s})
\end{array}
\right)
\label{eq:bidown_coreg}
\end{equation}
where, for each $k=1,\ldots, 6$, $\mbox{Cov}(w_{k}(\mathbf{s}),w_{k}(\mathbf{s^{\prime}}))=\exp(-\phi_{k} \| \mathbf{s} - \mathbf{s}^{\prime} \|)$, with $\phi_{k}$ decay parameter for $w_{k}(\mathbf{s})$.
As in the univariate case, the coregionalization matrix $\mathbf{A}$ determines, through $\mathbf{T}=\mathbf{A}\mathbf{A}^{\prime}$, the correlation between the six spatially-varying coefficients $\beta_{ij}(\mathbf{s})$.
Additionally, as a consequence of (\ref{eq:bidown_eq}), $\mathbf{A}$ induces a correlation structure on the bivariate random vector $\mathbf{Y}(\mathbf{s})$, i.e.
 \begin{equation}
{\boldsymbol\Sigma}_{Y(\mathbf{s}),Y(\mathbf{s}')} = \left[ \mathbf{I}_{2} \otimes (1 \; x_{1}(B) \; x_{2}(B) )^{\prime} \right]  \cdot \mathbf{A} {\boldsymbol\Sigma}_{w} \mathbf{A}^{\prime}
\cdot \left[ \mathbf{I}_{2} \otimes (1 \; x_{1}(B^{\prime}) \; x_{2}(B^{\prime})) \right]
\label{eq:gen_crosscov}
\end{equation}
where $B$ and $B^{\prime}$ are grid cells containing, respectively, $\mathbf{s}$ and $\mathbf{s}^{\prime}$, $\mathbf{I}_{2}$ is the identity matrix of order 2, and ${\boldsymbol\Sigma}_{w}$ is a $6\times 6$ diagonal matrix with
$i$-th element $\exp(-\phi_{i} \| \mathbf{s}-\mathbf{s}^{\prime} \|)$, $i=1,\ldots,6$.

Our bivariate downscaler model, though simple, is general and flexible.  The specification of the bias of the numerical model by means of spatially-varying coefficients recognizes that calibration
under the numerical model should be site-specific.  Again, as in the univariate case, a constant-mean bivariate downscaler can be obtained as a special case.  Moreover, permutation of the entries in the $\mathbf{\beta}(\mathbf{s})$ vector does not affect the prior.  The joint model still presents a $6 \times 6$ local covariance matrix modeled through its Cholesky decomposition.

Simplified versions of the general bivariate downscaler can be obtained by setting entries in the coregionalization matrix $\mathbf{A}$ equal to zero.\footnote{In principle, we could include all $21$ $A_{ij}$'s in the model and let the data suggest which are significant.  Instead, we have chosen to do model comparison between several models, each having plausible interpretations.}  For example, if the set of non-null entries in the
$\mathbf{A}$ matrix is given by $\left\{ A_{11}, A_{21}, A_{22}, A_{44}, A_{64}, A_{66} \right\}$ and we assume that $\beta_{12}\equiv 0$ and $\beta_{21}\equiv 0$, then our bivariate downscaler reduces to two independent univariate downscalers on ozone and PM$_{2.5}$, respectively.  This form of $\mathbf{A}$ implies that the spatially-varying coefficients are correlated within pollutants but not across pollutants.

A simpler version of the bivariate downscaler that requires only three non-null entries in the coregionalization matrix $\mathbf{A}$ but
still induces a correlation among the two components $Y_{1}(\mathbf{s})$ and $Y_{2}(\mathbf{s})$ of $\mathbf{Y}(\mathbf{s})$ is the model that assumes that only the local intercept adjustments,
$\beta_{10}(\mathbf{s})$ and $\beta_{20}(\mathbf{s})$, are non-null and correlated. In this case, the coregionalization matrix $\mathbf{A}$ has only three non-null entries, $\left\{ A_{11}, A_{41}, A_{44} \right\}$
and the covariance between the square root of the observed ozone concentration at $\mathbf{s}$ and the logarithm of the observed PM$_{2.5}$ concentration at $\mathbf{s}^{\prime}$
reduces simply to:
\begin{equation}
\mbox{Cov}(Y_{1}(\mathbf{s}),Y_{2}(\mathbf{s}^{\prime})) = A_{11}\cdot A_{41} \exp(-\phi_{1} \| \mathbf{s} - \mathbf{s}^{\prime} \|)
\label{eq:corr_bidown_mod1}
\end{equation}

A simple extension to this model can be achieved by maintaining the same correlation structure for the bivariate random vector $\mathbf{Y}(\mathbf{s})$, that is, by assuming again that $A_{41}$
is the only non-null off-diagonal element of the coregionalization matrix $\mathbf{A}$, but by postulating that the local adjustments to the coefficients of the two numerical model outputs,
$\beta_{11}(\mathbf{s})$, $\beta_{12}(\mathbf{s})$, $\beta_{21}(\mathbf{s})$, and $\beta_{22}(\mathbf{s})$, are non-null independent Gaussian processes with variances equal, respectively,
to $A_{22}^{2}$, $A_{33}^{2}$, $A_{55}^{2}$ and $A_{66}^{2}$. Then, the coregionalization matrix $\mathbf{A}$ corresponding to this model has seven non-null entries: $A_{11}$, $A_{22}$, $A_{33}$,
$A_{41}$, $A_{44}$, $A_{55}$, and $A_{66}$.

An extension of this last model that still involves fewer parameters than the general formulation for the bivariate downscaler (13 versus 21) is the model which assumes that only local adjustments of the form $\beta_{1j}(\mathbf{s})$ and $\beta_{2j}(\mathbf{s})$, for $j=0,1,2$, are correlated across pollutants,
while all the $\beta_{ij}(\mathbf{s})$'s specific to the same pollutant are correlated.
Such a model specification corresponds to the following coregionalization matrix:
\begin{equation}
\mathbf{A}=
\left(
\begin{array}{cccccc}
A_{11} & 0 & 0 & 0 & 0 & 0 \\
A_{21} & A_{22} & 0 & 0 & 0 & 0 \\
A_{31} & 0 & A_{33} & 0 & 0 & 0 \\
A_{41} & 0 & 0 & A_{44} & 0 & 0 \\
0  & A_{52} & 0 & A_{54} & A_{55} & 0 \\
0 & 0 & A_{63} & A_{64} & 0 & A_{66} \\
\end{array}
\right)
\label{eq:bidown_coreg_mod3}
\end{equation}
Other simplifications of the bivariate downscaler general model can be envisioned.

The specification of our Bayesian hierarchical bivariate downscaler model is completed with priors on the parameters.
We adopt an inverse gamma with large variance for the nugget variances,
$\tau^{2}_{1}$ and $\tau^{2}_{2}$; independent vague normals for the overall regression coefficients $\beta_{ij}$, $i=1,2$, $j=0,1,2$; lognormals with
vague standard deviations for the diagonal entries of the coregionalization matrix $\mathbf{A}$;
and vague normals for the off-diagonal elements of $\mathbf{A}$. Details on handling the decay parameters
$\phi_{k}$, $k=1,\ldots,6$, are provided in Section \ref{subsec:priors}.

\subsection{Space-time bivariate downscaler}\label{subsec:spacetime}
We now extend our general bivariate downscaler to accommodate data that has been collected over time.
Let  $t=1,\ldots,T$ denote the times
at which we have observations and numerical model outputs, and let $Y_{1}(\mathbf{s},t)$ and $Y_{2}(\mathbf{s},t)$ denote, respectively, the
square root and the logarithm of the observed ozone and PM$_{2.5}$ concentration at site $\mathbf{s}$ on day $t$. In an analogous way, let $x_{1}(B,t)$ and $x_{2}(B,t)$
indicate, respectively, the square root and the logarithm of the CMAQ output for ozone and PM$_{2.5}$ on day $t$ over the 12-km grid cell $B$.
Then, if $\mathbf{s}$ lies in grid cell $B$, the model for our bivariate downscaler becomes
\begin{eqnarray}
Y_{1}(\mathbf{s},t) & = & \tilde{\beta}_{10}(\mathbf{s},t) + \tilde{\beta}_{11}(\mathbf{s},t) x_{1}(B,t) + \tilde{\beta}_{12}(\mathbf{s},t) x_{2}(B,t) + \epsilon_{1}(\mathbf{s},t) \nonumber \\
Y_{2}(\mathbf{s},t) & = & \tilde{\beta}_{20}(\mathbf{s},t) + \tilde{\beta}_{21}(\mathbf{s},t) x_{1}(B,t) + \tilde{\beta}_{22}(\mathbf{s},t) x_{2}(B,t) + \epsilon_{2}(\mathbf{s},t)
\label{eq:bidown_eq_time}
\end{eqnarray}
where $\epsilon_{1}(\mathbf{s},t)$ and $\epsilon_{2}(\mathbf{s},t)$ are two white noise processes that follow independently a $N(0,\tau^{2}_{1})$ and a $N(0,\tau^{2}_{2})$ distribution.

As in the spatial setting, for each $i=1,2$ and $j=0,1,2$, we write:
\begin{equation}
\tilde{\beta}_{ij}(\mathbf{s},t) = \beta_{ij,t} + \beta_{ij}(\mathbf{s},t)
\label{eq:beta_time}
\end{equation} where, for each $t=1,\ldots, T$, $\beta_{ij}(\mathbf{s},t)$ are correlated Gaussian processes.

Following \cite{Berrocal&2009}, we can model the temporal dependence in the components of (\ref{eq:beta_time}) in several ways.
For example, for the $\beta_{ij,t}$'s, we can assume that they are either independent across time, i.e. $\beta_{ij,t} \stackrel{ind}{\sim} N(\mu_{ij},\sigma^{2}_{ij})$,
or, alternatively, that they evolve dynamically in time \cite{WestHarrison1999}, that is,
\begin{equation}
\beta_{ij,t} = \rho_{ij} \beta_{ij,t-1} + \eta_{ij,t}, \qquad \eta_{ij,t} \stackrel{ind}{\sim} N(0,\xi^{2}_{ij}), \qquad i=1,2; j=0,1,2
\label{eq:beta_dynam}
\end{equation}
and $\beta_{ij,0} \sim N(\mu_{ij,0}, \sigma^{2}_{ij,0})$.

Similar time-dependence can be imposed on the Gaussian processes $\beta_{ij}(\mathbf{s},t)$.
We can assume that the correlated Gaussian processes $\beta_{ij}(\mathbf{s},t)$ are of the form
\begin{equation}
\left(
\begin{array}{c}
\beta_{10}(\mathbf{s},t)  \\
\beta_{11}(\mathbf{s},t) \\
\beta_{12}(\mathbf{s},t)  \\
\beta_{20}(\mathbf{s},t) \\
\beta_{21}(\mathbf{s},t)  \\
\beta_{22}(\mathbf{s},t) \\
\end{array}
\right)
= \mathbf{A}
\left(
\begin{array}{c}
w_{1}(\mathbf{s},t) \\
w_{2}(\mathbf{s},t) \\
w_{3}(\mathbf{s},t) \\
w_{4}(\mathbf{s},t) \\
w_{5}(\mathbf{s},t) \\
w_{6}(\mathbf{s},t)
\end{array}
\right)
\label{eq:bidown_coreg_time}
\end{equation}
with $\mathbf{A}$ still a coregionalization matrix and with the underlying Gaussian processes, $w_{k}(\mathbf{s},t)$,
nested within time. Then, at each $t=1,\ldots,T$, the $w_{k}(\mathbf{s},t)$ are independent replicates of mean-zero unit-variance Gaussian processes with exponential correlation and decay parameters $\phi_{k,t}$ that can be either taken to be constant in time or independent across time.

In the second case, we can model the local adjustments, $\beta_{ij}(\mathbf{s},t)$, to evolve dynamically in time. As in \cite{Gelfand&2005}, for
each $i=1,2$ and $j=0,1,2$, we assume that
\begin{equation}
\beta_{ij}(\mathbf{s},t) = \gamma_{ij} \beta_{ij}(\mathbf{s},t-1) + \nu_{ij}(\mathbf{s},t)
\label{eq:beta_st_time}
\end{equation}
where the innovations $\nu_{ij}(\mathbf{s},t)$ are correlated Gaussian processes. In other words, in this second case,
the coregionalization (\ref{eq:bidown_coreg_time}) is specified on the $\nu_{ij}(\mathbf{s},t)$, rather than on the $\beta_{ij}(\mathbf{s},t)$, but it still employs the underlying Gaussian processes $w_{k}(\mathbf{s},t)$, which are defined as above. This dynamic model for the $\beta_{ij}(\mathbf{s},t)$ is completed by specifying as initial conditions that $\beta_{ij}(\mathbf{s},0)=0$ for $i=1,2$ and $j=0,1,2$.  In both cases we can be more general and assume that the coregionalization matrix in (\ref{eq:bidown_coreg_time}) is indexed by time, with entries independent across time.  Furthermore, if $t$ is on a continuous domain, then the $w_{k}(\mathbf{s},t)$ can be independent space time processes.

The two different time dependence structures for the overall regression coefficients, $\beta_{ij,t}$, and for their local adjustments, $\beta_{ij}(\mathbf{s},t)$, can be combined together in four ways, yielding four
models that correspond to different assumptions on the way the overall and local performance and calibration of the numerical model changes over time.

\subsection{The general multivariate downscaler}\label{subsec:multivariate}
Extension to a multivariate downscaler given observational data and numerical model outputs for $p$ random variables is evident.
As in the bivariate case, we start with the static formulation of the model and then we extend it to the spatio-temporal setting.

Let $Y_{i}(\mathbf{s})$, $i=1,\ldots,p$ be the observed data for the $i$-th variable at a site $\mathbf{s}$ in the spatial domain $\mathcal{S}$ and let $x_{i}(B)$, $i=1,\ldots,p$, be the numerical model output for the
$i$-th variable over grid cell $B$. Then, again we associate to each site $\mathbf{s}$ the grid cell $B$ in which $\mathbf{s}$ lies and we
relate the observational data and the numerical model output as follows:
\begin{equation}
Y_{i}(\mathbf{s}) = \tilde{\beta}_{i0}(\mathbf{s}) + \sum_{j=1}^{p} \tilde{\beta}_{ij}(\mathbf{s}) x_{j}(B) + \epsilon_{i}(\mathbf{s}) \qquad \epsilon_{i}(\mathbf{s}) \stackrel{iid}{\sim} N(0,\tau^{2}_{i})
\label{eq:multidown_eq}
\end{equation}
for $i=1,\ldots,p$.

We decompose each of the $p \cdot (p+1)$ terms $\tilde{\beta}_{ij}(\mathbf{s})$, $i=1,\ldots,p$, $j=0,1,\ldots,p$ in the sum of an overall term and a local adjustment:
$\tilde{\beta}_{ij}(\mathbf{s})  =  \beta_{ij} + \beta_{ij}(\mathbf{s})$, and we model the $\beta_{ij}(\mathbf{s})$'s as correlated mean-zero Gaussian processes with exponential
covariance structures using again the method of coregionalization.
In the general $p$-dimensional multivariate downscaler, the coregionalization matrix $\mathbf{A}$ is a $p(p+1) \times p(p+1)$ matrix.  Specifications of (\ref{eq:multidown_eq}) similar to the above can be considered by simply setting to zero entries of the $\mathbf{A}$ matrix, thus inducing simplifications
on both the correlation structure of the multivariate random vector $\mathbf{Y} = \left\{ (Y_{i}(\mathbf{s}))_{i=1,\ldots,p} : \mathbf{s} \in \mathcal{S} \right\}$ and
on the covariance structure of the single $Y_{i}(\mathbf{s})$, $i=1\ldots,p$.

To complete the specification of the model, we place standard priors on the model parameters: thus, we specify $ p(p+1)$
vague normals for the overall regression coefficients $\beta_{ij}$; $p$ inverse gammas with large variances for the nugget variances $\tau^{2}_{i}$, $i=1\ldots,p$; log-normals
with large variances for the diagonal entries
$A_{kk}$, $k=1,\ldots,p(p+1)$, of the coregionalization matrix $\mathbf{A}$; and vague normals for the off-diagonal entries of $\mathbf{A}$.

If the data on the $p$ random variables has been collected not only over space, but also over time, then the multivariate general downscaler would be extended to a spatio-temporal setting in an obvious way.
Let $Y_{i}(\mathbf{s},t)$, $i=1,\ldots,p$ be the observed data for the $i$-th random variable at site $\mathbf{s}$ at time $t$, $t=1,
\ldots,T$, and let $x_{i}(B,t)$ be the model output for the $i$-th variable over grid cell $B$ at time $t$. Then, for $i=1,\ldots,p$, we postulate the following relationship between observational data and numerical model output:
\begin{equation}
Y_{i}(\mathbf{s},t) = \tilde{\beta}_{i0}(\mathbf{s},t) + \sum_{j=1}^{p} \tilde{\beta}_{ij}(\mathbf{s},t) x_{j}(B,t) + \epsilon_{i}(\mathbf{s},t) \qquad \epsilon_{i}(\mathbf{s},t) \stackrel{iid}{\sim} N(0,\tau^{2}_{i})
\label{eq:multidown_eq_time}
\end{equation}
where $\mathbf{s}$ lies in grid cell $B$.

For each $i=1,\ldots,p$ and $j=0,1,\ldots,p$, we write: $\tilde{\beta}_{ij}(\mathbf{s},t)=\beta_{ij,t} + \beta_{ij}(\mathbf{s},t)$, and we model the temporal structure in the $\beta_{ij,t}$ and in the
$\beta_{ij}(\mathbf{s},t)$, following what we have presented in Section \ref{subsec:spacetime} for the bivariate downscaler. Thus, we assume that the $\beta_{ij,t}$'s are either nested within time or dynamic
in time, and, similarly, we postulate that the $\beta_{ij}(\mathbf{s},t)$ are either independent replicates over time or are dynamic in time.

\section{Model fitting}\label{sec:fitting}

\subsection{Priors}\label{subsec:priors}
We have already mentioned briefly in Sections \ref{subsec:bivariate} and \ref{subsec:multivariate} the form of the prior distributions for the parameters of the bivariate and multivariate downscaler both in their static and spatio-temporal formulations.
Here, we turn to the spatial decay parameters $\phi_{k}$ of the latent Gaussian processes $w_{k}(\mathbf{s})$ used in the coregionalization.  Discrete uniform priors facilitate model fitting.
Previous experience with such priors has always resulted in
posterior distributions placing highest probability on a value which is, essentially, the Restricted Maximum Likelihood (REML; \cite{PattersonThompson1971,Harville1977}) estimate. Hence, in what follows, we propose to keep the spatial decay parameters fixed and equal to values that are determined by a sensitivity analysis.

More specifically, consider the simplified bivariate downscaler that uses only the two latent Gaussian processes $w_{1}(\mathbf{s})$ and $w_{4}(\mathbf{s})$ and is obtained when the coregionalization
matrix $\mathbf{A}$ has only three non-null entries, $A_{11}, A_{41}$, and $A_{44}$. The decay parameters $\phi_{1}$ and $\phi_{4}$ determine the way the spatial correlation decays with distance in $w_{1}(\mathbf{s})$ and $w_{4}(\mathbf{s})$, respectively, and, by consequence, in $Y_{1}(\mathbf{s})$ and $Y_{2}(\mathbf{s})$. In fact, the covariance structure of $Y_{2}(\mathbf{s})$ induced by such simplified version of the bivariate downscaler is given by the sum of two exponential covariance functions with decay parameters, respectively, equal to $\phi_{1}$ and $\phi_{4}$, and a diagonal matrix with entries all equal to the nugget variance $\tau^{2}_{2}$.

To obtain a rough estimate of the magnitude of the decay parameters $\phi_{1}$ and $\phi_{4}$, we have proceeded as follows. We have considered the monitoring data for the square root of ozone and the logarithm of PM$_{2.5}$ for a given day $t$. We have modeled the monitoring data for both variables as Gaussian processes with an unknown mean, respectively, $\mu_{Y_{1,t}}$ and $\mu_{Y_{2,t}}$, and with an exponential covariance function plus nugget effect, with decay parameters $\phi_{Y_{1,t}}$ and $\phi_{Y_{2,t}}$, and nugget variance, $\tau^{2}_{Y_{1,t}}$ and $\tau^{2}_{Y_{2,t}}$. We have estimated the parameters via Restricted Maximum Likelihood (REML; \cite{PattersonThompson1971, Harville1977} as implemented in the \texttt{geoR} package of \texttt{R}. We have repeated this operation for each day in the period June 1-- September 30, 2002 and have obtained daily estimates of $\phi_{Y_{1,t}}$ and $\phi_{Y_{2,t}}$. Since histograms of those daily estimates exhibited long right tails, we have summarized those distributions by taking the median of the daily estimates, which yielded, respectively, $0.0016$ and $0.00125$, corresponding to ranges of, respectively, 1875 and 2400 km.
We have then performed a sensitivity analysis to determine how the predictive performance of a downscaler model is affected by the magnitude of the decay parameter. Therefore, keeping the decay parameters $\phi_{1}$ and $\phi_{4}$ fixed and equal to the estimated medians of $\phi_{Y_{1,t}}$ and $\phi_{Y_{2,t}}$, we have fit the spatio-temporal version of the bivariate downscaler that models the overall terms $\beta_{ij,t}$, $i=1,2$, $j=0,1,2$ and the local adjustments to the overall intercepts, $\beta_{i0}(\mathbf{s},t)$, $i=1,2$ as independent in time (see Section 7.2 in the Supplemental material online). For each day, we have predicted levels of ozone and PM$_{2.5}$ concentration at the validation sites $\mathbf{s}_{0}$ by sampling from the posterior predictive distributions, $f( \mathbf{Y}(\mathbf{s}_{0}) | \left\{ \mathbf{Y} (\mathbf{s}) \right\} , \left\{ \mathbf{x} (B) \right\} )$ and subsequently backtransforming the prediction to the original scale.  We have then compared the predictions with the observations by computing the predictive mean square error (PMSE), predictive mean absolute error (PMAE), empirical coverage of the 95$\%$ predictive interval and width of the 95$\%$ predictive interval. More specifically, denoting with $Y^{\star}(\mathbf{s},t)$ the backtransformed data (recall we used the square root and the logarithm transform on ozone and PM$_{2.5}$, respectively), we have computed the PMSE as follows:
\begin{equation}
\mbox{PMSE}=\frac{1}{n_{v}} \sum_{t=1}^{T} \sum_{r=1}^{V} \left( \widehat{Y^{\star}}(\mathbf{s}_{r},t)-Y^{\star}(\mathbf{s}_{r},t) \right)^{2} \cdot \mathbf{I}(Y^{\star}(\mathbf{s}_{r},t))
\label{eq:pmse}
\end{equation}
where $n_{v}$ denotes the total number of observations in the validation dataset (6530 for ozone or 2559 for PM$_{2.5}$ as noted in Section \ref{sec:data}), $\mathbf{s}_{r}$ is the $r$-th site in the validation set containing a total of $V$ sites, $\widehat{Y^{\star}}(\mathbf{s}_{r},t)$ denotes the posterior mean of the predictive distribution at site $\mathbf{s}_{r}$ at time $t$ on the original scale, and $\mathbf{I}(Y^{\star}(\mathbf{s}_{r},t))$ is equal to $1$ if $Y^{\star}(\mathbf{s}_{r},t)$ is observed and $0$ otherwise. An analogous definition holds for the PMAE, with $\widehat{Y^{\star}}(\mathbf{s}_{r},t)$ now referring to the posterior median of the predictive distributions, again on the original scale.  We have generated predictions and computed the summary statistics mentioned above four times, each time keeping the decay parameters fixed and setting them equal, respectively, to the estimated medians of $\phi_{Y_{1,t}}$ and $\phi_{Y_{2,t}}$, multiplied by 10 and 100, and divided by 10 and 100.

To compare the performance of the bivariate downscaler to that of a univariate downscaler, we have performed the same sensitivity analysis also for the univariate downscaler. Thus, keeping the decay parameters for ozone and PM$_{2.5}$ fixed and equal to the values considered for the bivariate downscaler, we have fit two spatio-temporal univariate independent downscalers for ozone and particulate matter, where the only non-null local adjustment term was the adjustment to the overall intercept, and where both the overall regression terms and the spatially varying coefficients were nested within time. Then, as in the bivariate downscaler case, we have predicted ozone and PM$_{2.5}$ concentration at the validation sites and we have assessed the performance of the predictions using the same summary statistics mentioned above.

\begin{table}[!hb]
\begin{center}
\caption{Predictive Mean Square Error, Predictive Mean Absolute Error, empirical coverage and width of the 95$\%$ predictive interval for $O_3$ for different values of the decay parameter
under different models.}
\vspace{0.5cm}
\begin{tabular}{|c|c||c|c|c|c|c|}
\hline
                              &                                     &    \multicolumn{5}{|c|}{Decay parameter}  \\
                                   \cline{3-7}
                              &  Summary                  &                       &                       &                          &                     &            \\
Model                   &   statistic                    &     0.16e-4     &   0.16e-3     &      0.16e-2      &     0.16e-1  & 0.16  \\
\hline \hline
                                 &   PMSE                   &   55.54 &   55.94  &  55.92   &  58.13  &    95.17 \\ \cline{2-7}
   Independent       &  PMAE                   &  5.54   &  5.56   &  5.56   & 5.68  &  7.39  \\   \cline{2-7}
  downscaler          & Emp. cov.             & 93.2$\%$  &  92.0$\%$ & 92.9$\%$  & 91.7$\%$  & 92.9$\%$ \\  \cline{2-7}
                                  &  Width                  &   27.33  & 26.30   &  27.34  & 27.19 &    35.78 \\
 \hline                         \hline
                               &   PMSE                   &   53.21   &   51.64   & 52.78   & 55.24   & 91.42   \\  \cline{2-7}
  Bivariate             &  PMAE                   &   5.45   &   5.36   &  5.40  &   5.54  &   7.27  \\  \cline{2-7}
 downscaler         & Emp. cov.             &   92.6$\%$  & 92.3$\%$    &  93.0$\%$  & 92.7$\%$  & 93.1$\%$   \\  \cline{2-7}
                               &  Width                  &  26.67   & 26.25   & 26.79   & 27.77  &  34.67   \\
 \hline               \hline
                                                  &   PMSE                   &  144.58   &   83.28  &   57.16  &  77.19    &  218.03  \\  \cline{2-7}
Ordinary                                   &  PMAE                   &  9.31 & 6.89   & 5.62  & 6.51   & 11.52  \\   \cline{2-7}
kriging                & Emp. cov.             & 59.5$\%$  & 76.6$\%$ & 88.2$\%$ & 92.8$\%$ & 78.0$\%$  \\    \cline{2-7}
                                  &  Width                  &  19.10   &  19.73  &  22.26  &  30.26    &  34.69  \\
\hline
\end{tabular}
\label{tbl:pmse_ozone_sens}
\end{center}
\end{table}

We have carried out a similar sensitivity analysis also for ordinary kriging \cite{Cressie1993,ChilesDelfiner1999}. Therefore, for both ozone and PM$_{2.5}$, for each day, using the same range of values for the decay parameters used in the experiments with the univariate and bivariate downscalers, we have kriged the observed data, on the transformed scale, to the validation sites, and we have subsequently backtransformed the predictions. Finally, we have compared the predictions with the observations using the same summary statistics that we have employed for both downscalers.

\begin{table}[!hb]
\begin{center}
\caption{Predictive Mean Square Error, Predictive Mean Absolute Error, empirical coverage and width of the 95$\%$ predictive interval for PM$_{2.5}$ for different values of the decay parameter
under different models.}
\vspace{0.5cm}
\begin{tabular}{|c|c||c|c|c|c|c|}
\hline
                             &                                     &         \multicolumn{5}{|c|}{Decay parameter}  \\
                                   \cline{3-7}
                              &  Summary                  &                       &                       &                          &                     &            \\
Model                   &   statistic                    &     0.125e-4     &   0.125e-3     &      0.125e-2     &     0.125e-1  & 0.125  \\
\hline \hline
                                   &   PMSE                   &   13.46  &   13.27  &  12.95   &  16.58 &   36.43 \\ \cline{2-7}
  Independent         &  PMAE                   &   2.43   &   2.42   &   2.41   &  2.65  &   3.83  \\   \cline{2-7}
  downscaler           & Emp. cov.             & 93.0$\%$  &   91.8$\%$  &  92.7$\%$  & 91.6$\%$   &  92.7$\%$ \\  \cline{2-7}
                                  &  Width                  &    15.01   & 14.29   &  14.79 &  15.05 &    20.17 \\
 \hline      \hline
                                   &   PMSE                   &   11.98    &  11.78   & 11.69   & 14.43   & 32.55   \\  \cline{2-7}
  Bivariate         &  PMAE                   &   2.33   &   2.31   &  2.29  &   2.49  &   3.55  \\  \cline{2-7}
  downscaler           & Emp. cov.             &   92.7$\%$   & 93.6$\%$    &   94.0$\%$ &  93.2$\%$ & 93.5$\%$    \\  \cline{2-7}
                                  &  Width                  &  14.02   &  14.33     & 14.53   & 14.87   &  18.63   \\
 \hline      \hline
                                   &   PMSE                   &  43.89   &   21.99    &   16.19 &  25.65    &  64.17  \\   \cline{2-7}
                                  &  PMAE                   &   4.37 & 3.06   & 2.54  & 3.11   &  5.37  \\    \cline{2-7}
  Kriging                  & Emp. cov.             & 52.2$\%$  &  72.5$\%$  &  89.7$\%$  & 94.8$\%$  & 89.9$\%$ \\  \cline{2-7}
                                  &  Width                  &  7.32     &   8.89   & 14.06   &  22.61   &  25.73 \\
\hline
\end{tabular}
\label{tbl:pmse_pm25_sens}
\end{center}
\end{table}

Table \ref{tbl:pmse_ozone_sens} and Table \ref{tbl:pmse_pm25_sens} report, respectively, the Predictive Mean Square Error (PMSE), Predictive Mean Absolute Error (PMAE), the empirical coverage and the width of the nominal $95 \%$ predictive interval for ozone and PM$_{2.5}$ for the two space-time downscalers and for ordinary kriging. In both tables, the results corresponding to predictions obtained when the models have been
fitted using values of the decay parameters equal to the medians of the daily estimates of $\phi_{Y_{1,t}}$ and $\phi_{Y_{2,t}}$ are reported in the middle column.  From both tables, it is clear that when the decay parameter is one and, especially, two orders of magnitude larger, the quality of the predictions obtained using both downscalers deteriorates noticeably in terms of PMSE and PMAE.
Also, the predictive intervals are wider in this case, an indication that there is increased variability in the predictions. Misspecifying the magnitude of the decay parameters on the small side affects the quality of the predictions less noticeably.
So, altogether, the sensitivity analysis has shown that both the univariate and the bivariate downscaler and ordinary kriging yield predictions that have overall best validation statistics when the decay parameters are equal to the median of the daily estimates of $\phi_{Y_{1,t}}$ and $\phi_{Y_{2,t}}$ and so, in the sequel, we fix them at these values.

\subsection{Handling misalignment}\label{subsec:misalignment}
As noted in Section \ref{sec:data}, not all sites reporting ozone concentration measure PM$_{2.5}$ and vice versa. This creates a spatial misalignment in the data that we handle
through the latent Gaussian processes $w_{k}(\mathbf{s})$ involved in the coregionalization, after having appropriately permuted and partitioned the data vector. Here, we illustrate how to handle the
spatial misalignment in the static setting. Extension to the spatio-temporal case is straightforward.

Let $\mathcal{S}_{t}$ be the set of $n_{t}$ sites reporting measurements of ozone and/or PM$_{2.5}$ concentration on day $t$. We can decompose
$\mathcal{S}_{t}$ as the union of three disjoint sets, $\mathcal{S}_{\mbox{\scriptsize{Both}},t}$, $\mathcal{S}_{O,t}$, and $\mathcal{S}_{PM,t}$:
\begin{equation}
\mathcal{S}_{t} = \mathcal{S}_{\mbox{\scriptsize{Both}},t} \cup \mathcal{S}_{O,t} \cup \mathcal{S}_{PM,t}
\label{eq:decomp}
\end{equation}
where the first set includes all sites $\mathbf{s}\in \mathcal{S}$ in which both ozone and PM$_{2.5}$ have been measured on day $t$, the second contains all the sites in which only ozone
was measured on day $t$, and the last includes all sites $\mathbf{s}$ where only PM$_{2.5}$ has been measured on day $t$.

Following the decomposition in (\ref{eq:decomp}), we reorder and partition the random vectors $\mathbf{Y}=(Y_{1}(\mathbf{s}),Y_{2}(\mathbf{s}))_{\mathbf{s}\in \mathcal{S}_{t}}$ and
$\mathbf{w}_{k}=(\mathbf{w}_{k}(\mathbf{s}))_{\mathbf{s}\in \mathcal{S}_{t}}$ in the following way: $\mathbf{Y}=(\mathbf{Y}_{\mbox{\scriptsize{Both}}},\mathbf{Y}_{O},\mathbf{Y}_{PM})$ and $\mathbf{w}_{k}=(\mathbf{w}_{k,\mbox{\scriptsize{Both}}},\mathbf{w}_{k,O},\mathbf{w}_{k,PM})$
where $\mathbf{Y}_{\mbox{\scriptsize{Both}}}=(Y_{1}(\mathbf{s}),Y_{2}(\mathbf{s}))_{\mathbf{s} \in \mathcal{S}_{\mbox{\scriptsize{Both}},t}}$ and analogous definitions hold for $\mathbf{Y}_{O}$, $\mathbf{Y}_{PM}$, $\mathbf{w}_{k,\mbox{\scriptsize{Both}}}$,
$\mathbf{w}_{k,O}$, and $\mathbf{w}_{k,PM}$.

Then it is clear that both components $\mathbf{Y}_{1,\mbox{\scriptsize{Both}}}=(Y_{1}(\mathbf{s}))_{\mathbf{s} \in \mathcal{S}_{\mbox{\scriptsize{Both}},t}}$ and
$\mathbf{Y}_{2,\mbox{\scriptsize{Both}}}=(Y_{2}(\mathbf{s}))_{\mathbf{s} \in \mathcal{S}_{\mbox{\scriptsize{Both}},t}}$
have non-missing values and both contribute to the log-likelihood (\ref{eq:loglik_data}), while for $\mathbf{s} \in \mathcal{S}_{O,t}$ and $\mathbf{s} \in \mathcal{S}_{PM,t}$ only one component,
respectively, $\mathbf{Y}_{1,O}$ and $\mathbf{Y}_{2,PM}$, has non-missing values and contributes to the log-likelihood.

The latent Gaussian processes $w_{k}(\mathbf{s})$ are defined on the entire spatial domain $\mathcal{S}$. To obtain samples from the posterior distribution of $w_{k}(\mathbf{s})$, within
each MCMC iteration, we draw samples from the full conditionals by proceeding in the following way.
If $\Theta$ denotes the collection of all model parameters (thus, $\Theta=(\tau^{2}_{1},\tau^{2}_{2},\left\{ A_{kl} \right\}_{k,l=1,\ldots,6; l \leq k}, \left\{ \beta_{ij} \right\}_{i=1,2; j=0,1,2})$),
for each $k=1,\ldots,6$, we:
\begin{enumerate}[label=\textbf{\roman{*}.}, ref=(\roman{*})]
\item sample $\mathbf{w}_{k,\mbox{\scriptsize{Both}}}$ from the full conditional $\pi (\mathbf{w}_{k,\mbox{\scriptsize{Both}}} | \mathbf{w}_{k,O},\mathbf{w}_{k,PM}, \mathbf{Y}_{\mbox{\scriptsize{Both}}}, \Theta)$;
\item sample $\mathbf{w}_{k,O}$ from the full conditional $\pi (\mathbf{w}_{k,O} | \mathbf{w}_{k,\mbox{\scriptsize{Both}}},\mathbf{w}_{k,PM}, \mathbf{Y}_{1,O}, \Theta)$;
\item sample $\mathbf{w}_{k,PM}$ from the full conditional $\pi (\mathbf{w}_{k,PM} | \mathbf{w}_{k,\mbox{\scriptsize{Both}}},\mathbf{w}_{k,O}, \mathbf{Y}_{2,PM}, \Theta)$;
\item for each $\mathbf{s}\in \mathcal{S} \setminus \mathcal{S}_{t}$, sample ${w}_{k}(\mathbf{s})$ from the conditional distribution $\pi (w_{k}(\mathbf{s}) | \mathbf{w}_{k,\mbox{\scriptsize{Both}}},
\mathbf{w}_{k,O} , \mathbf{w}_{k,PM}, \Theta)$
\end{enumerate}

By repeating steps \textbf{i}-\textbf{iv} within each MCMC iteration, we obtain samples from the posterior distribution of ${w}_{k}(\mathbf{s})$ for $\mathbf{s}\in \mathcal{S}$, thus
solving the problem of spatial misalignment between ozone and PM$_{2.5}$.

\subsection{Model selection}\label{subsec:modelsel}
Using results from an analysis on the predictive performance of the univariate spatio-temporal downscaler for ozone concentration \cite{Berrocal&2009},
we consider only the spatio-temporal versions of the bivariate downscaler with
time-varying parameters independent across time. More specifically, we have fitted the spatio-temporal versions of the four bivariate downscalers that we
have presented in Section \ref{subsec:bivariate}, that is:
\begin{enumerate}[label=\textbf{\roman{*}.}, ref=(\roman{*})]
\item the bivariate downscaler equivalent to two independent univariate downscalers, obtained when the coregionalization matrix $\mathbf{A}$ has
as non-null entries only \\ $\left\{ A_{11},A_{21},A_{22},A_{44},A_{64},A_{66} \right\}$ ;
\item the bivariate downscaler with only $\left\{ A_{11}, A_{41}, A_{44} \right\}$ as non-null entries in the coregionalization matrix $\mathbf{A}$;
\item the bivariate downscaler with $\left\{ A_{11}, A_{22}, A_{33}, A_{41}, A_{44}, A_{55}, A_{66} \right\}$ as non-null entries in the coregionalization matrix $\mathbf{A}$; and
\item the bivariate downscaler with coregionalization matrix $\mathbf{A}$ given by (\ref{eq:bidown_coreg_mod3})
\end{enumerate}

For each of these models, we have examined their out-of-sample predictive performance, by predicting daily
concentrations of ozone and PM$_{2.5}$ at the 65 validation sites described in Section \ref{sec:data} and shown in Figure~\ref{fig:studyarea}.
These predictions are compared with the observed
ozone and PM$_{2.5}$ levels at the validation sites during the period June 1-September 30, 2002.
Note that, for each day, predictions of ozone and PM$_{2.5}$ at a validation site $\mathbf{s}_{0}$ are obtained by
sampling from the posterior predictive distribution $f(\mathbf{Y}(\mathbf{s}_{0},t) |
\left\{ \mathbf{Y}(\mathbf{s},t) \right\}, \left\{ \mathbf{x}(B,t) \right\})$
and then by subsequently transforming them back to the original scale.

To quantify the predictive performance of each bivariate downscaler we have utilized the same validation statistics used in our sensitivity analysis presented in Section \ref{subsec:priors}. We also employ two proper scoring rules: (i) the Continuous Ranked Probability Score (CRPS; \cite{GneitingRaftery2007}) defined as:
$$
\mbox{CRPS}(F,y)=\int  (F(z) - \mathbf{1}_{ \left\{ z \geq y \right\} })^{2} d\;z  
$$
where $F$ is the cumulative predictive distribution function, $y$ is the observation that materializes, $\mathbf{1}$ is the Heaviside function, that is equal to $1$ if $z$ is greater than $y$ and $0$ otherwise, and (ii) the Interval Score \cite{GneitingRaftery2007}, defined for a $(1-\alpha)\cdot 100\%$ predictive interval with lower bound $l$ and upper bound $u$ as:
$$
\mbox{IS}(l; u; y)=\left[ (u-l) + \frac{2}{\alpha} (l-y) \mathbf{1}_{\left\{ y < l  \right\}} + \frac{2}{\alpha}
(u-y) \mathbf{1}_{\left\{ u < y \right\}} \right] 
$$
where $y$ denotes again the observed value.

So, for each model, we have computed the Predictive Mean Square Error (PMSE), Predictive Mean Absolute Error (PMAE), the empirical coverage of the 95$\%$ predictive interval, and the width of the 95$\%$ predictive interval to obtain an indication not only of the average bias and errors in the predictions, but also of the level of uncertainty in the predictions. Additionally, the CRPS, being a strictly proper scoring rule, provides a simultaneous assessment of the calibration and sharpness of the posterior predictive distribution whereas the Interval Score rewards predictive distributions with narrower predictive intervals while imposing a penalty for observations lying outside the predictive interval.

Finally, to determine the improvement in the predictions of ozone and PM$_{2.5}$ obtained from using our downscaling approach with respect to non-model based techniques, we have
compared the downscaler with ordinary kriging and cokriging \cite{Cressie1993,ChilesDelfiner1999}. For kriging, using as
decay parameters for ozone and PM$_{2.5}$ concentration, respectively, 0.0016 and 0.00125, for each day, we have kriged, separately, the observed daily concentrations of ozone and PM$_{2.5}$ at the test sites to the validation sites, working on the transformed scale and then backtransforming the predictions. For cokriging, we have exploited the information contained in the copollutant and, for each day, we have cokriged the observed daily concentrations of ozone, first, and PM$_{2.5}$, afterwards, to the validation sites. Note that in performing cokriging, we had to define, not only a covariance function for both ozone and PM$_{2.5}$, but also a cross-covariance function. For both pollutants, as covariance function we used again the exponential covariance function with decay parameters, respectively, equal to 0.0016 and 0.00125.
For cross-covariance function, instead, we used the cross-covariance function induced by the second version of the bivariate downscaler listed above, explicitly given in (\ref{eq:corr_bidown_mod1}).
The parameters for such cross-covariance function were set to be equal to the posterior means of $A_{11}, A_{41}$ and $A_{44}$ obtained from fitting the bivariate downscaler model in \textbf{ii}.
Finally, predictive intervals and uncertainty for kriging and cokriging are based on usual formulas for the kriging and cokriging variance \cite{ChilesDelfiner1999}.

\section{Data Analysis}\label{sec:analysis}

We first present validation results for the different models. Since we did not find a significant improvement in the predictions of ozone and PM$_{2.5}$ obtained from
versions \textbf{iii} and \textbf{iv} of the bivariate downscaler, we show only results for the versions, \textbf{i} and \textbf{ii}.  To distinguish between the two, we will refer to the first as the independent downscaler, while the second will be called the bivariate downscaler.
Comparing the performance of the two models allows us to establish the gain in predictive performance that can be ascribed to explicitly accounting for the correlation between the two pollutants.

From Section \ref{subsec:modelsel}, for each of the models considered, we have computed the predictive mean square, the predictive mean absolute error, the empirical coverage, the width of the 95$\%$ predictive intervals, the Continuous Ranked Probability Score and the Interval Score, averaging across sites and days. In order to determine whether distance from the closest monitoring sites has an effect on the quality of the predictions, we have divided the validation sites into two groups: those that have a distance from the closest monitoring test site of less or equal than 40 kilometers, and those that are more than 40 kilometers far apart from the closest monitoring test sites.
For each of the two groups, consisting, respectively, of 13 and 52 sites, we have computed the same validation statistics mentioned above.

Table \ref{tbl:result}, Table \ref{tbl:result_le40}, and Table \ref{tbl:result_gt40} report all these summary statistics, respectively, for all the validation sites, for the validation sites that are less or equal than 40 km
distant from the closest monitoring test sites, and for the validation sites that are more than 40 km away from the closest monitoring test site.

From all tables, it is clear that the bivariate downscaler yields predictions of both ozone and PM$_{2.5}$ concentrations that are less biased than those obtained using
the univariate downscaler. Additionally, the bivariate downscaler has a lower average CRPS score, indicating that its predictive distribution is sharper and better calibrated than the one corresponding to the univariate downscaler.  In terms of width and empirical coverage of the 95$\%$ predictive intervals, both downscalers perform similarly and both have empirical coverage close to nominal. However, the combined assesment of both properties, which is provided by the interval score, favors again the bivariate downscaler over the univariate downscaler.
A similar trend in predictive performance can be observed for kriging and cokriging: exploiting the correlation between ozone and particulate matter yield predictions that are less biased than those obtained by not accounting for it. Additionally, kriging always underestimates uncertainty in the predictions, producing predictive intervals that are too narrow and, thus, do not have the nominal coverage.  This is also reflected in the CRPS and the Interval Scores: kriging always has a lower CRPS score than cokriging probably due to the smaller widths of the predictive intervals, which in turn, failing to achieve the nominal coverage are penalized in terms of Interval Score.

Comparing the bivariate downscaler with cokriging, we can see that the two methods perform equally well in terms of Interval Score; however the former outperforms the latter in terms of PMSE, PMAE, and CRPS for both ozone and PM$_{2.5}$.
This follows from the fact that the bivariate downscaler not only models the correlation between ozone and PM$_{2.5}$ but
also uses the information contained in the numerical model output, while cokriging does not take advantage of this additional information when predicting levels of ozone and PM at the validation sites.
In addition, the bivariate downscaler predicts levels of PM$_{2.5}$ better than cokriging, likely because of the sampling frequency of PM$_{2.5}$
which renders it rather difficult to interpolate to the entire spatial domain, especially when the number of sites with observations is fairly low. Note that validation
of the out-of-sample predictive performance of  both downscalers and kriging/cokriging for PM$_{2.5}$ occurs mostly every three days due to
the sampling frequency of PM$_{2.5}$. In days with fewer observations from monitoring sites, this can diminish predictive gain associated with
both downscalers relative to spatial interpolation techniques based solely on monitoring data.

The separation of validation sites into two groups based on their distance to the closest monitoring test site, does not reveal a difference in terms of the predictive performance of the various models. The bivariate downscaler still outperforms all the other models. However, when considering sites that are less than 40 kilometers far from the closest monitoring test site, the bivariate downscaler and cokriging
predict ozone concentration equally well. This is not true for particulate matter, nor it is true when we consider validation sites that are more than 40 kilometers away from the closest monitoring test site. For this group of sites, for both ozone and PM$_{2.5}$, the predictions obtained using the bivariate downscaler are superior to the predictions obtained using any other method.  Counterintuitively, for all methods, the quality of the predictions are better at validation sites that are farther from the closest monitoring test sites
than at validation sites that are closer to the monitoring test sites.

\begin{table}[!ht]
\begin{center}
\caption{Predictive Mean Square Error, Predictive Mean Absolute Error, empirical coverage, width of the 95$\%$ predictive
interval, Continuos Ranked Probability Score and Interval Score  for both pollutants for the different models, kriging and cokriging at all the 65 validation sites.}
\vspace{0.5cm}
\begin{tabular}{|c||c||c||c|c||c|c|}
\hline \hline
                             &              &                      &    Independent                &     Bivariate          &                         &                        \\
Pollutant             &              &     CMAQ     &     downscaler                  &      downscaler     &     Kriging      &   Cokriging    \\
\hline \hline
                            &   PMSE &  122.1  &   55.9     &           52.8          &     57.2   &    53.4  \\ \cline{2-7}
                            &   PMAE  &  8.5  &   5.6    &            5.4   &    5.6     &  5.4  \\   \cline{2-7}
 Ozone               &  Emp. cov.  &   --  &  92.9$\%$   &   93.0$\%$    &   88.2$\%$ &   94.3$\%$  \\   \cline{2-7}
                            &  Width         &  --  &  27.3&     26.8   &    22.3  &   27.9  \\   \cline{2-7}
                            &   CRPS      & 8.5  & 4.0 &  3.9  &  4.1  &  5.2  \\ \cline{2-7}
                            & Interval   &          &        &         &          &          \\
                            & score      &  --     &  40.0 &  38.9 &  43.0  &  37.1   \\
\hline \hline
                            &   PMSE &   63.8   & 13.0    &   11.7   &     16.2 &    13.9  \\   \cline{2-7}
                            &   PMAE  &  5.1     &   2.4    &  2.3    &    2.5   &  2.4 \\    \cline{2-7}
PM$_{2.5}$       &  Emp. cov.  &   --   &  92.7$\%$   &   94.0$\%$  &   89.7$\%$      & 92.6$\%$  \\    \cline{2-7}
                            &  Width         &  --     &  14.8          &    14.6        &   14.1     &  12.8  \\  \cline{2-7}
                            & CRPS         &  5.1 &  1.8 &  1.7 & 1.9  & 2.3 \\ \cline{2-7}
                            & Interval   &          &        &         &          &          \\
                            & score      &  --     &  19.2 &  17.8 & 21.1 & 19.6  \\
 \hline \hline
\end{tabular}
\label{tbl:result}
\end{center}
\end{table}

\begin{table}[!ht]
\begin{center}
\caption{Predictive Mean Square Error, Predictive Mean Absolute Error, empirical coverage and width of the 95$\%$ predictive
interval, Continuous Ranked Probability Score and Interval Score for both pollutants for the different models, kriging and cokriging at the 13 validation sites that
are at no more than 40 km of distance from the closest monitoring test site.}
\vspace{0.5cm}
\begin{tabular}{|c||c||c||c|c||c|c|}
\hline \hline
                             &              &                   &   Independent                &     Bivariate          &                         &                        \\
Pollutant             &              &   CMAQ    &     downscaler                  &      downscaler     &     Kriging      &   Cokriging    \\
\hline \hline
                            &   PMSE &  124.4  &  57.4    &           54.1          &     57.9  &    54.2 \\ \cline{2-7}
                            &   PMAE  &  8.5  &  5.6   &            5.4   &    5.6   &  5.5 \\   \cline{2-7}
 Ozone               &  Emp. cov.  &  --  &   92.7$\%$ &   92.7$\%$ &   88.4$\%$  &   94.3$\%$ \\   \cline{2-7}
                            &  Width         &  -- &   27.3 &     26.7   &    22.2 &   27.9 \\     \cline{2-7}
                           & CRPS         & 8.5  &  4.1 & 4.0  &  4.1 &  5.2 \\ \cline{2-7}
                            & Interval   &          &        &         &          &          \\
                            & score      &  --     &  41.0 &  39.9 &  43.4 &  37.7 \\
\hline \hline
                            &   PMSE &   67.9 &  13.3    &   12.1    &     16.9  &    14.5 \\   \cline{2-7}
                            &   PMAE  &  5.2 &  2.4   &  2.3  &    2.6  &  2.4 \\    \cline{2-7}
PM$_{2.5}$       &  Emp. cov.  &   --   &  92.8$\%$  &   94.4$\%$  &   89.9$\%$    & 92.9$\%$ \\    \cline{2-7}
                            &  Width         & --  &  13.8          &    13.4        &   12.1  &  12.9  \\  \cline{2-7}
                           & CRPS         &  5.2 &  1.8 &  1.7 & 1.9  & 2.3 \\ \cline{2-7}
                            & Interval   &          &        &         &          &          \\
                            & score      &  --     &   19.9 &  18.3  & 22.0 &  18.8  \\
 \hline \hline
\end{tabular}
\label{tbl:result_le40}
\end{center}
\end{table}

\begin{table}[!ht]
\begin{center}
\caption{Predictive Mean Square Error, Predictive Mean Absolute Error, empirical coverage and width of the 95$\%$ predictive
interval, Continuous Ranked Probability Score and Interval Score  for both pollutants for the different models, kriging and cokriging at the 52 validation sites that are at more than 40 km of
distance from the closest monitoring test site.}
\vspace{0.5cm}
\begin{tabular}{|c||c||c||c|c||c|c|}
\hline \hline
                             &              &                   &          Independent                &     Bivariate          &                         &                        \\
Pollutant             &              &    CMAQ   &  downscaler                  &      downscaler     &     Kriging      &   Cokriging    \\
\hline \hline
                            &   PMSE &   113.1 &  50.1    &            47.4          &     54.4  &     50.1 \\ \cline{2-7}
                            &   PMAE  &   8.2 & 5.3   &             5.2   &    5.6   &   5.3 \\   \cline{2-7}
 Ozone               &  Emp. cov.  &  -- &   93.7$\%$  &   94.3$\%$  &   87.5$\%$  &   94.5$\%$ \\   \cline{2-7}
                            &  Width         &   -- &   27.1 &      26.5   &    22.2 &   27.8 \\    \cline{2-7}
                           & CRPS         & 8.2  &   3.9 &   3.8 &  4.1 & 5.1 \\ \cline{2-7}
                            & Interval   &          &        &         &          &          \\
                            & score      &  --     &  36.1 & 34.7   & 41.4 &  35.0 \\
\hline \hline
                            &   PMSE &   48.5  &   11.6   &    10.0    &     13.5  &     11.4 \\   \cline{2-7}
                            &   PMAE  &  4.6 &   2.4   &  2.2  &     2.4  &  2.3 \\    \cline{2-7}
PM$_{2.5}$       &  Emp. cov.  &   --   &  92.3$\%$  &   92.9$\%$  &   88.8$\%$     & 91.4$\%$ \\    \cline{2-7}
                            &  Width     & --    &  13.1          &     12.8        &    11.6   &  12.3  \\   \cline{2-7}
                           & CRPS         & 4.6  &  1.7 &  1.6 & 1.8  &  2.2 \\ \cline{2-7}
                            & Interval   &          &        &         &          &          \\
                            & score      &  --     &  16.6 & 15.7  & 17.6 & 15.6  \\
 \hline \hline
\end{tabular}
\label{tbl:result_gt40}
\end{center}
\end{table}

\begin{figure}[!hp]
\centering
\begin{tabular}{cc}
\includegraphics[scale=0.3,angle=0]{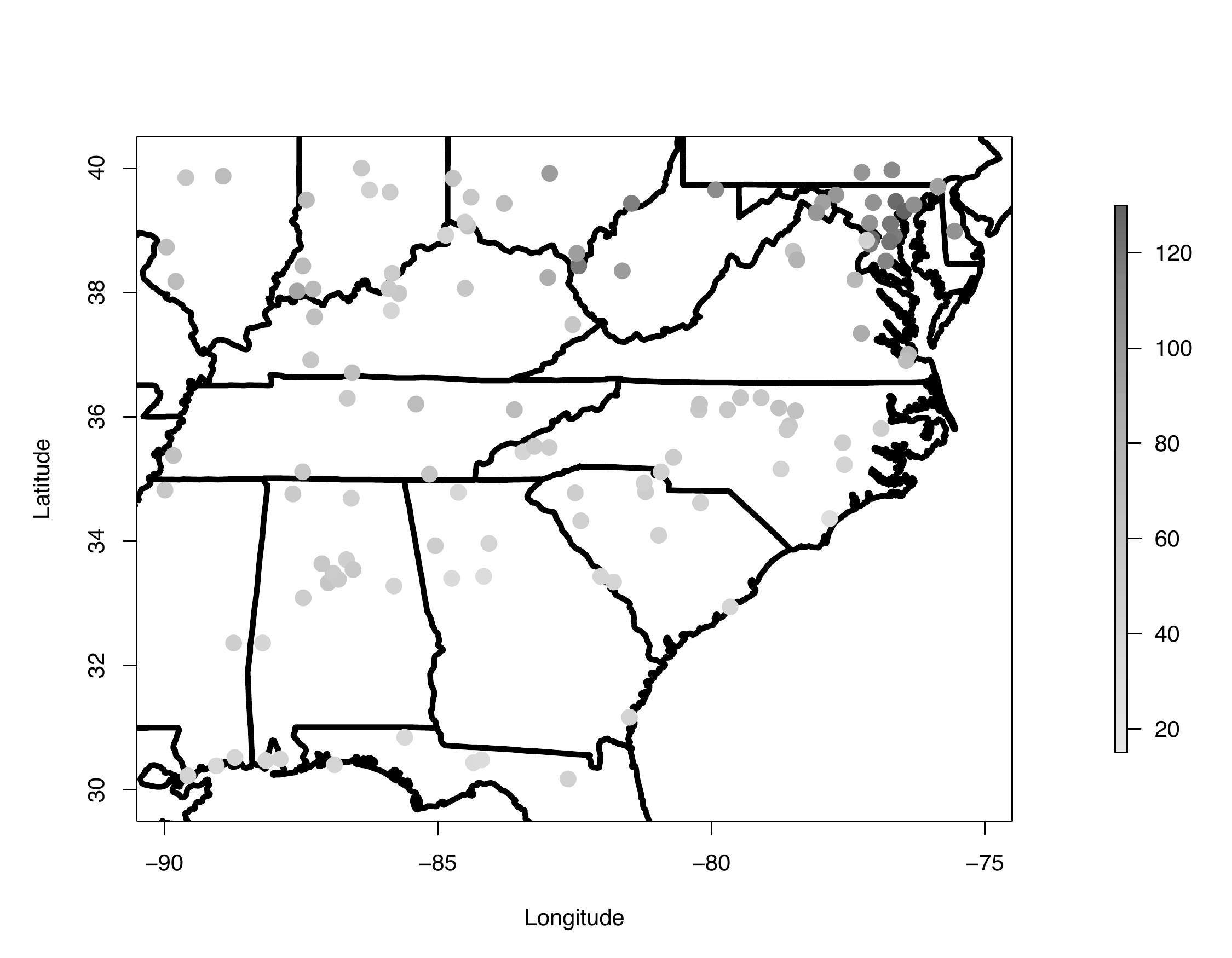}
&
\includegraphics[scale=0.3,angle=0]{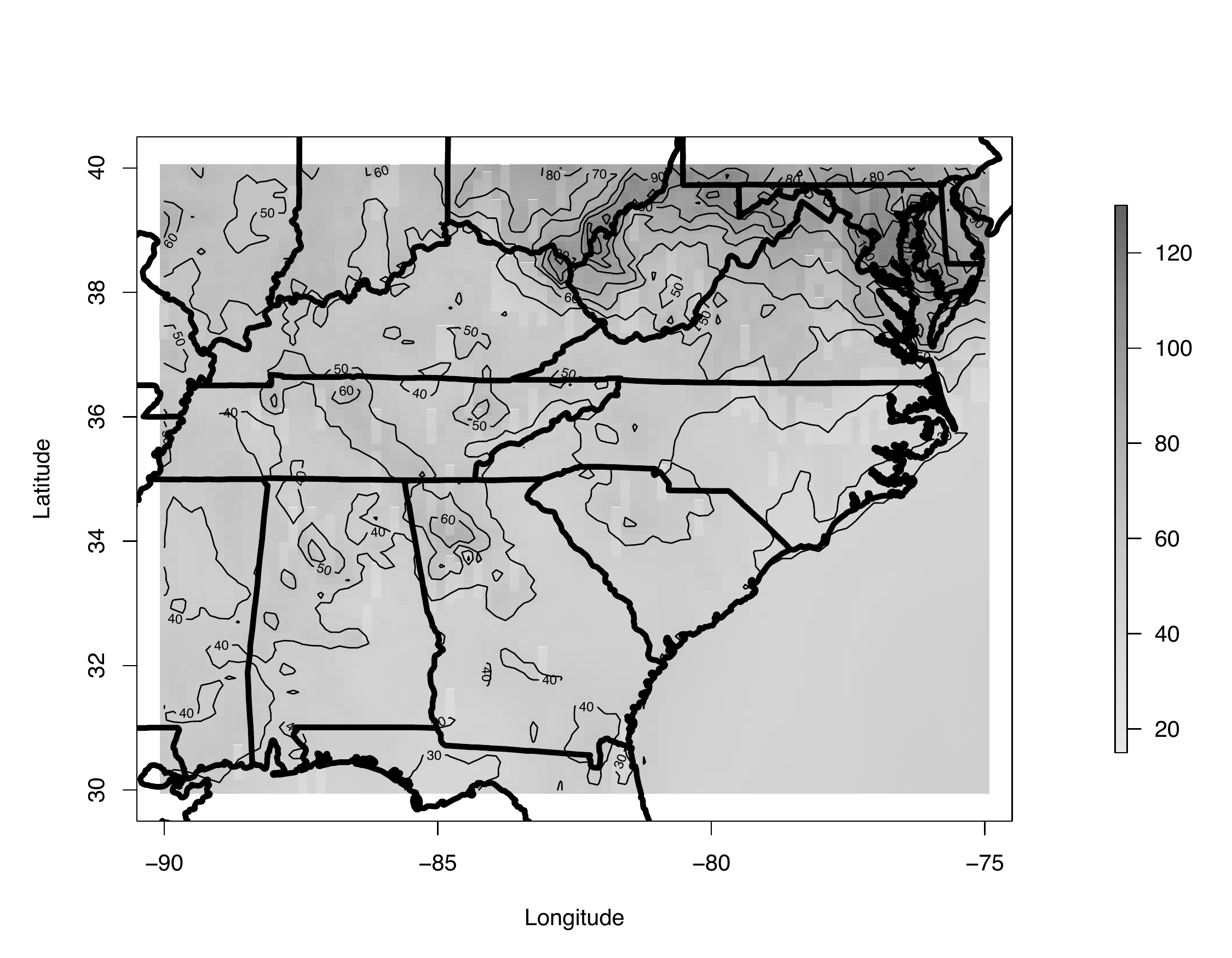}
\\
(a) & (b) \\
\\
\includegraphics[scale=0.3,angle=0]{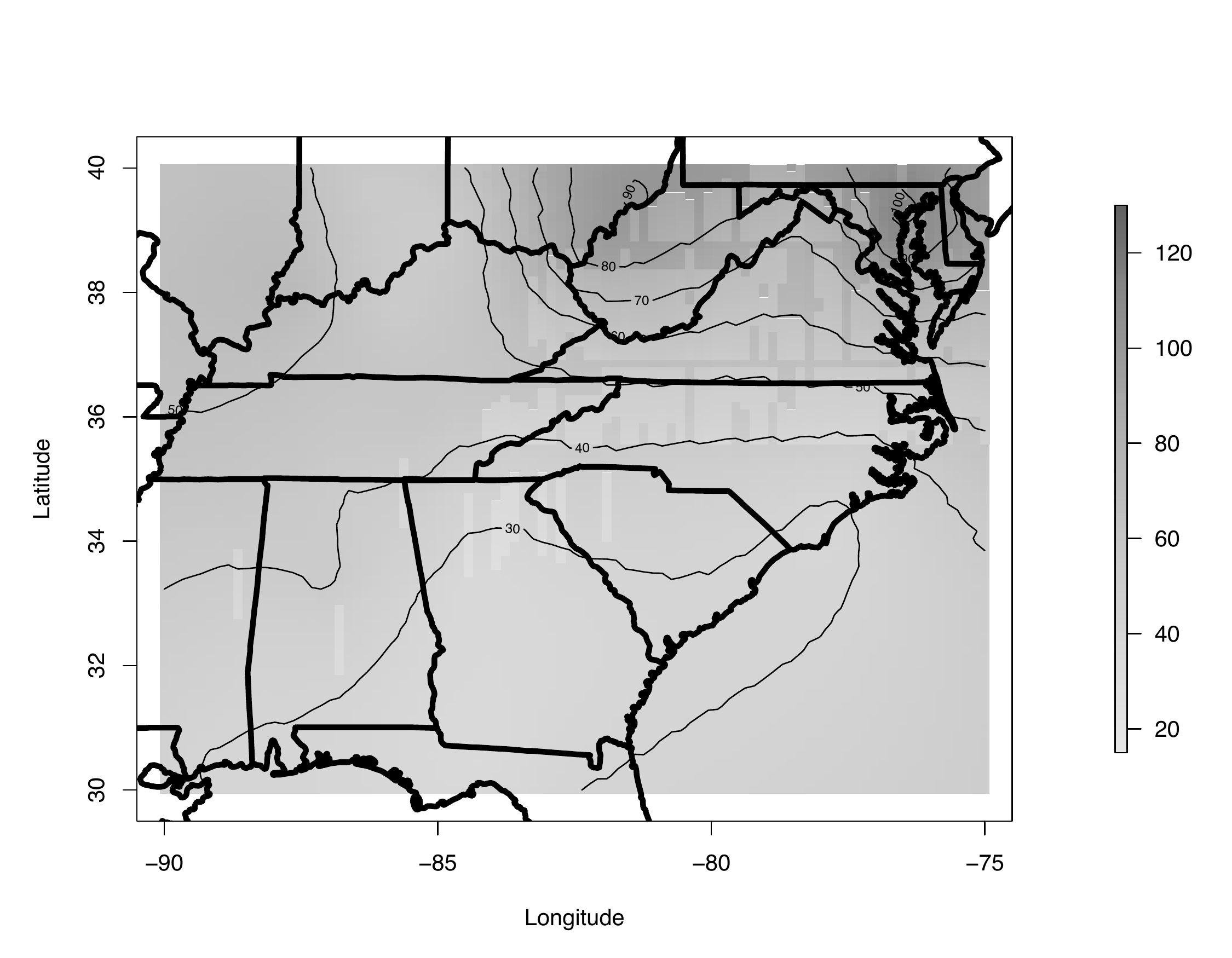}
&
\includegraphics[scale=0.3,angle=0]{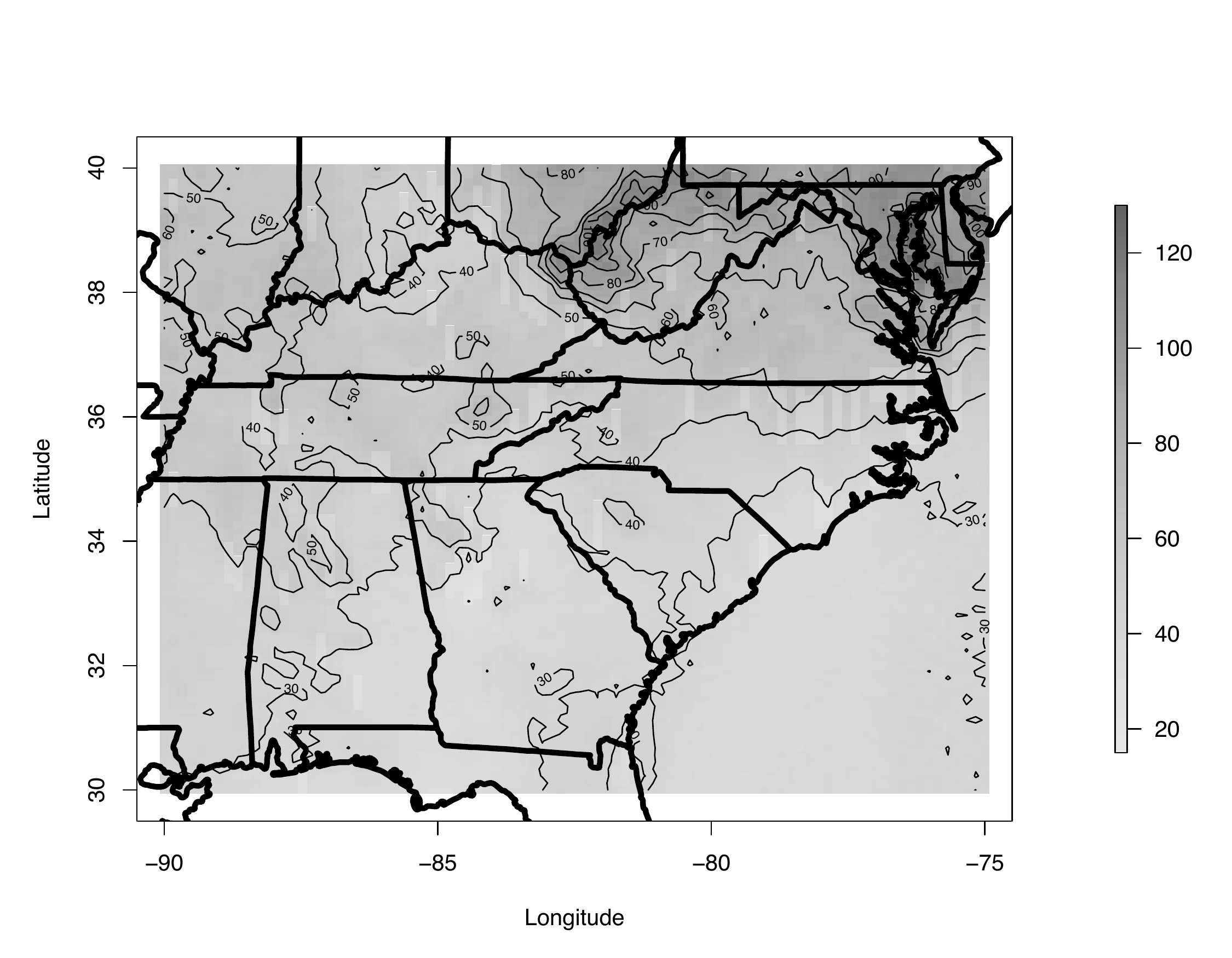}
\\
(c) & (d) \\
\end{tabular}
\caption{(a) Observed ozone on June 25, 2002; (b) Predicted ozone as obtained from CMAQ
on June 25, 2002; (c) Predicted ozone as obtained via kriging for June 25, 2002; (d) Predicted ozone as obtained from the bivariate downscaler
model for June 25, 2002.
\label{fig:ozonejun25}}
\end{figure}

\begin{figure}[!hp]
\centering
\begin{tabular}{cc}
\includegraphics[scale=0.3,angle=0]{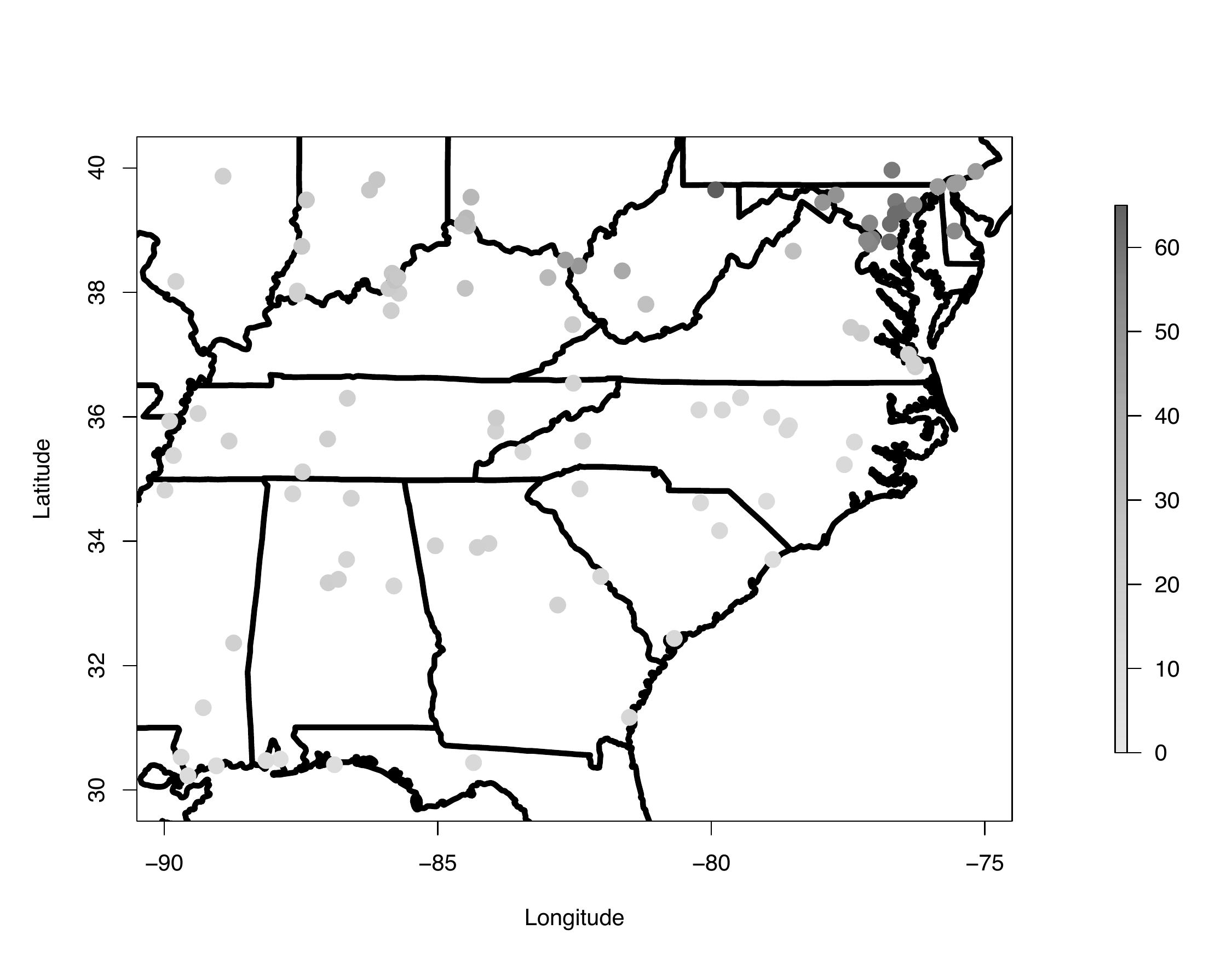}
&
\includegraphics[scale=0.3,angle=0]{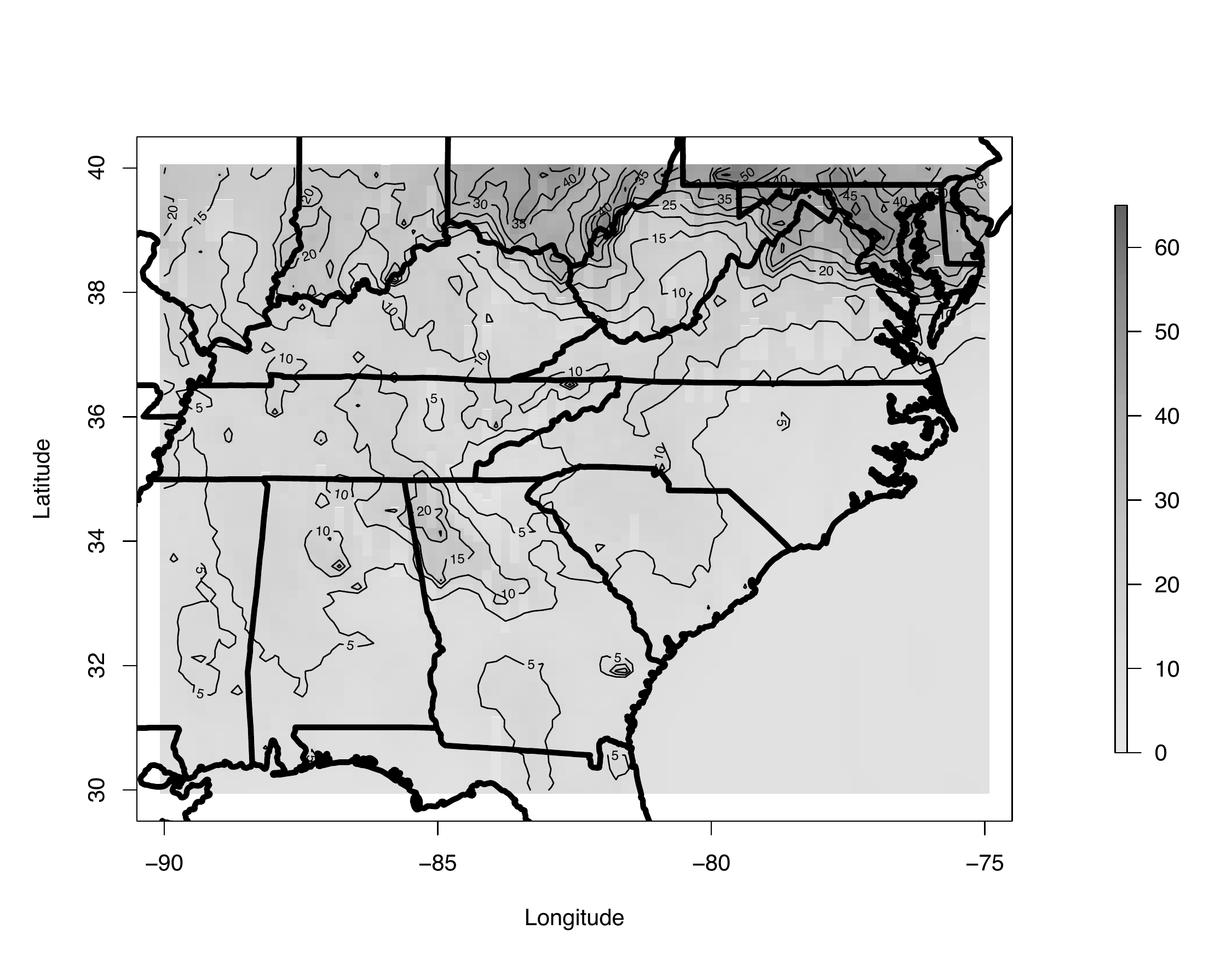}
 \\  
(a) & (b) \\
\\
\includegraphics[scale=0.3,angle=0]{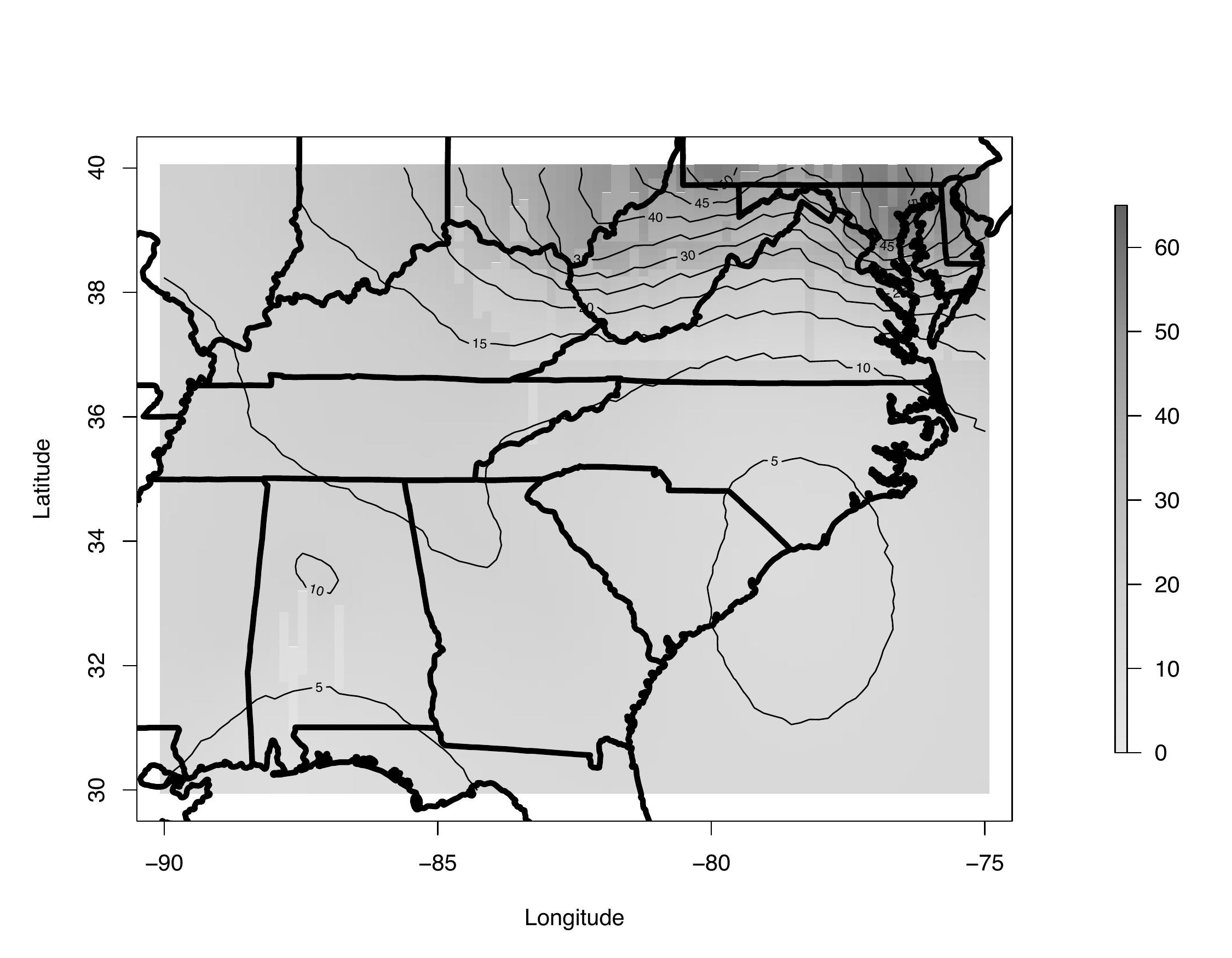} 
&
\includegraphics[scale=0.3,angle=0]{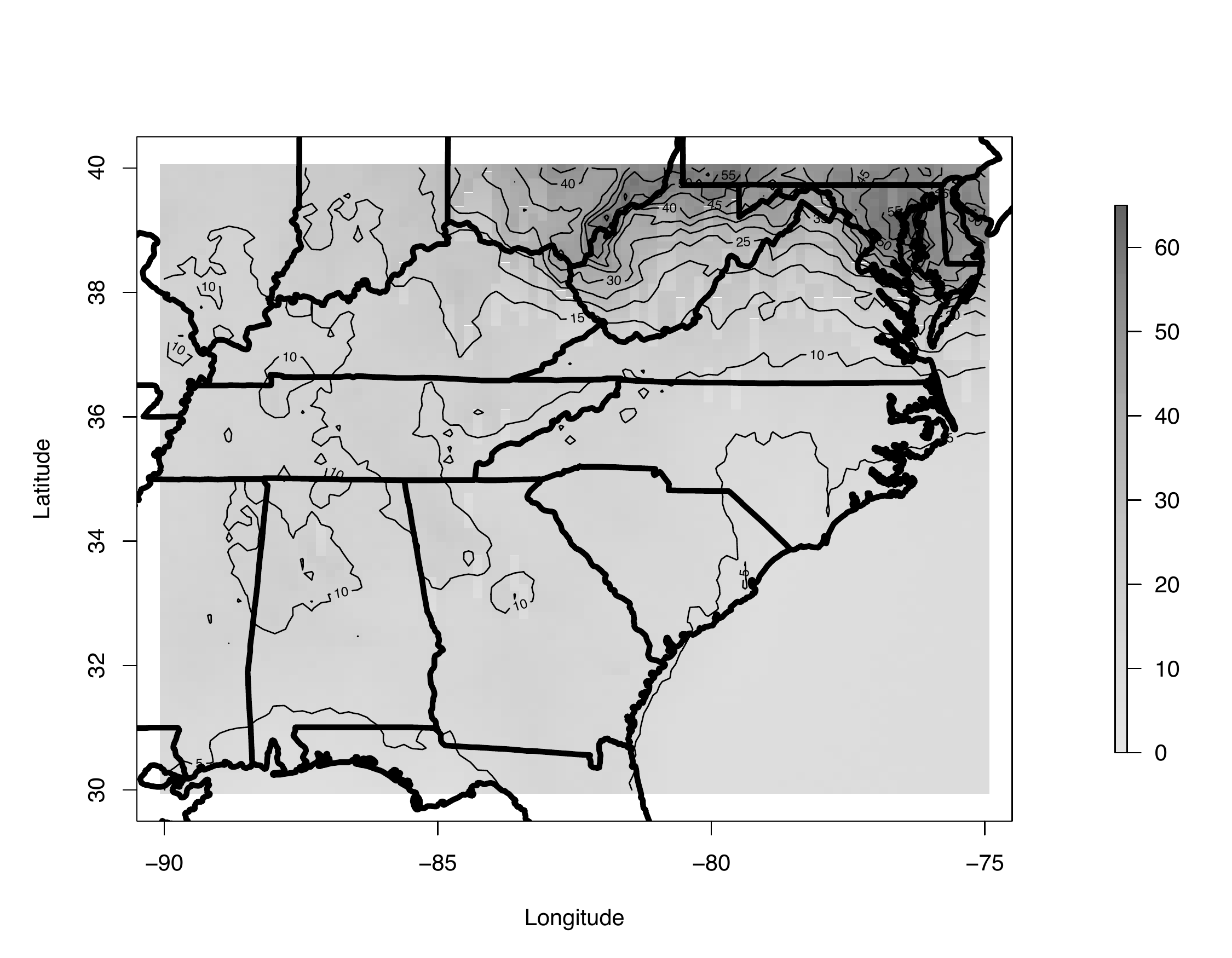}
\\
(c) & (d) \\
\end{tabular}
\caption{(a) Observed PM$_{2.5}$ on June 25, 2002; (b) Predicted PM$_{2.5}$ as obtained from CMAQ
on June 25, 2002; (c) Predicted PM$_{2.5}$ as obtained via kriging for June 25, 2002; (d) Predicted PM$_{2.5}$ as obtained from
the bivariate downscaler model for June 25, 2002.
\label{fig:pmjun25}}
\end{figure}

Our bivariate downscaler model, as well as the univariate version, can also be used to generate predictive surfaces for both ozone and PM$_{2.5}$ concentration.
Figure~\ref{fig:ozonejun25} and Figure~\ref{fig:pmjun25} display, in panels (d), predictive surfaces of ozone and PM$_{2.5}$ concentration for June 25, 2002
obtained using our bivariate downscaler model. Each panel displays the posterior predictive mean of ozone and PM$_{2.5}$ concentration obtained by sampling
from the posterior predictive distribution at each site $\mathbf{s}_{0}$ on the CMAQ grid covering the study region.
As a comparison, Figure~\ref{fig:ozonejun25} and Figure~\ref{fig:pmjun25} present also, in panel (c), the predictive surfaces obtained by simply kriging
the observed concentrations of ozone and PM$_{2.5}$ to the 10,504 CMAQ grid cells covering
the region. Finally, panels (a) and (b) show respectively the observed and the CMAQ predicted concentrations for both pollutants.

As both figures show, the predictive surfaces obtained using kriging are very smooth for both pollutants, while the surfaces obtained using our bivariate downscaler show more \emph{texture} which is expected to provide improved prediction. In particular, our bivariate downscaler method generates predictive surfaces that correct the CMAQ predictions for local bias, and display
a gradient that reproduces some of the features of the CMAQ surfaces while accounting for the spatial gradient in the observations.
For example, Figure~\ref{fig:ozonejun25}(d) predicts lower level of ozone than CMAQ in the South-Eastern corner of map, while Figure~\ref{fig:pmjun25}(d)
pushes up the level of PM$_{2.5}$ predicted by CMAQ in the same region.

\begin{figure}[!hp]
\vspace{-2cm}
\centering
\begin{tabular}{cc}
\includegraphics[scale=0.3,angle=0]{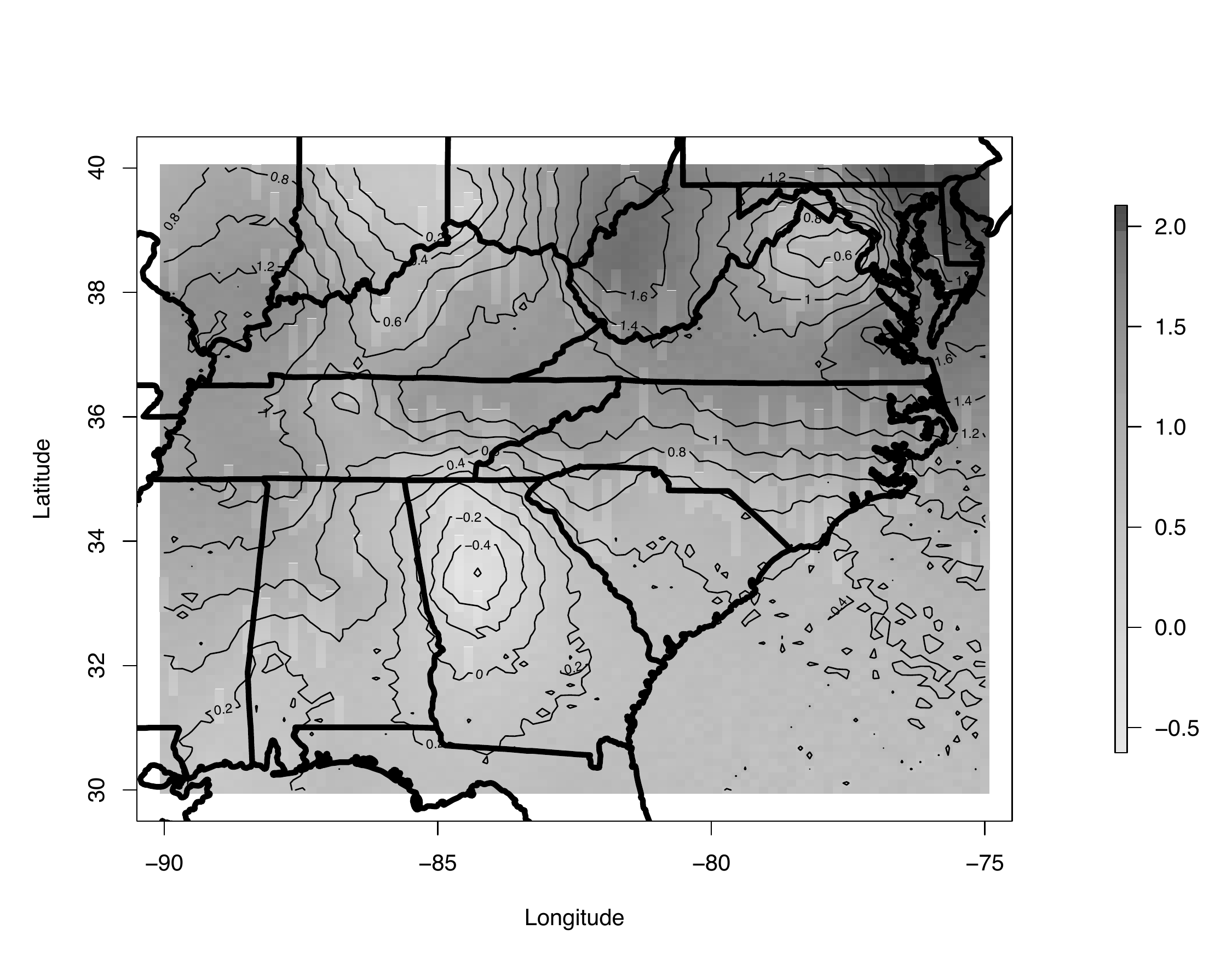} &
\includegraphics[scale=0.3,angle=0]{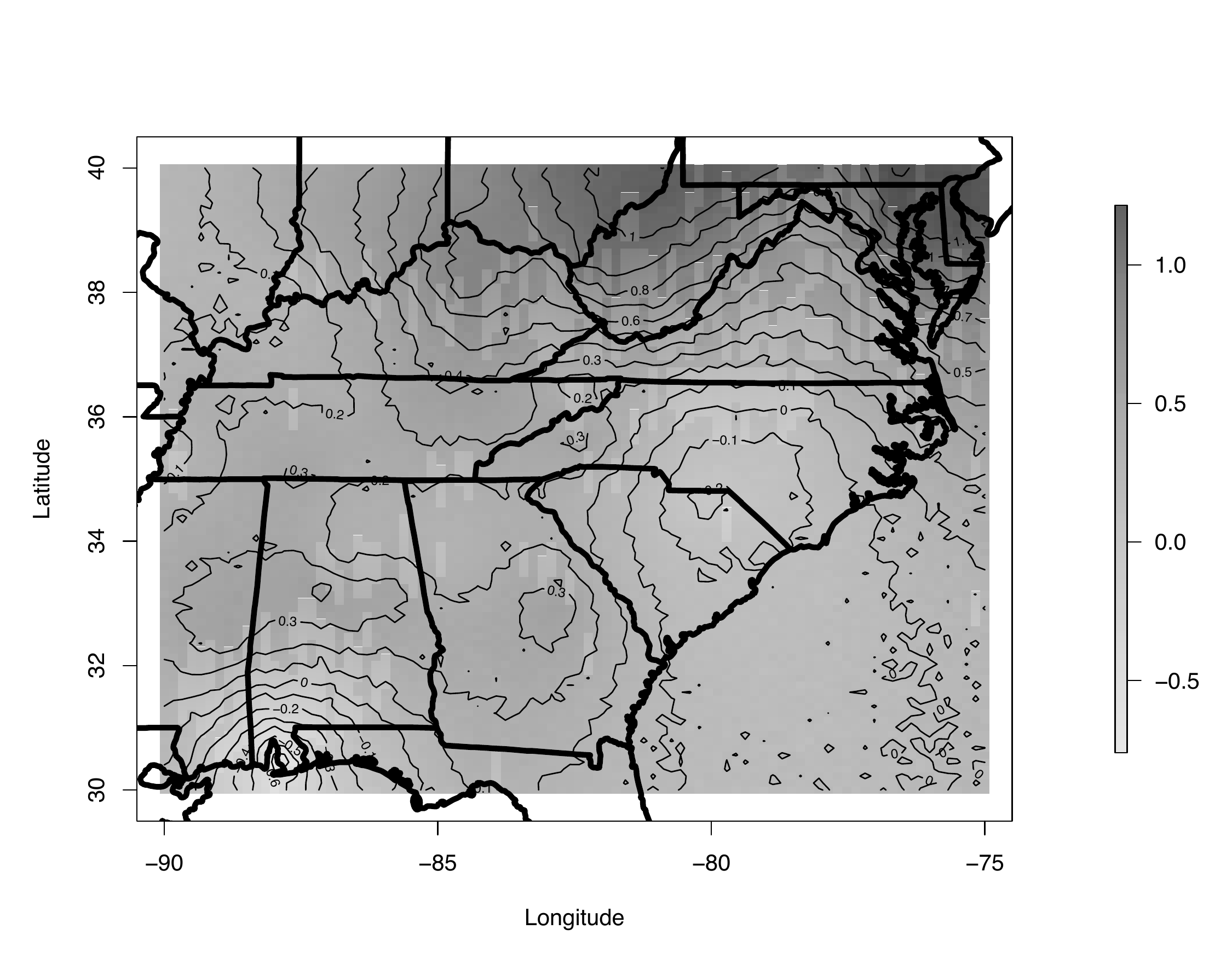}
\\
(a) & (b) \\
\end{tabular}
\caption{Posterior predictive mean of: (a) $\beta_{10}(\mathbf{s},t)$ for June 25, 2002; (b) $\beta_{20}(\mathbf{s},t)$ for June 25, 2002 as obtained from the bivariate downscaler
model.
\label{fig:beta_jun25}}
\end{figure}

In order to visually quantify the local biases of the CMAQ predicted concentrations of ozone and PM$_{2.5}$ for June 25, 2002, Figure~\ref{fig:beta_jun25}
displays spatial maps of the posterior predictive means of $\beta_{10}(\mathbf{s})$ and $\beta_{20}(\mathbf{s})$. Since the bivariate downscaler models has
been developed on the transformed scale, immediate conversion of these maps into calibration for the numerical model is not possible. However,
they do display the spatial variability in the additive bias of the numerical model output as well as identify areas where the numerical model either underpredicts
or overpredicts a pollutant concentration.

\begin{figure}[!hp]
\centering
\begin{tabular}{ccc}
\includegraphics[scale=0.2,angle=0]{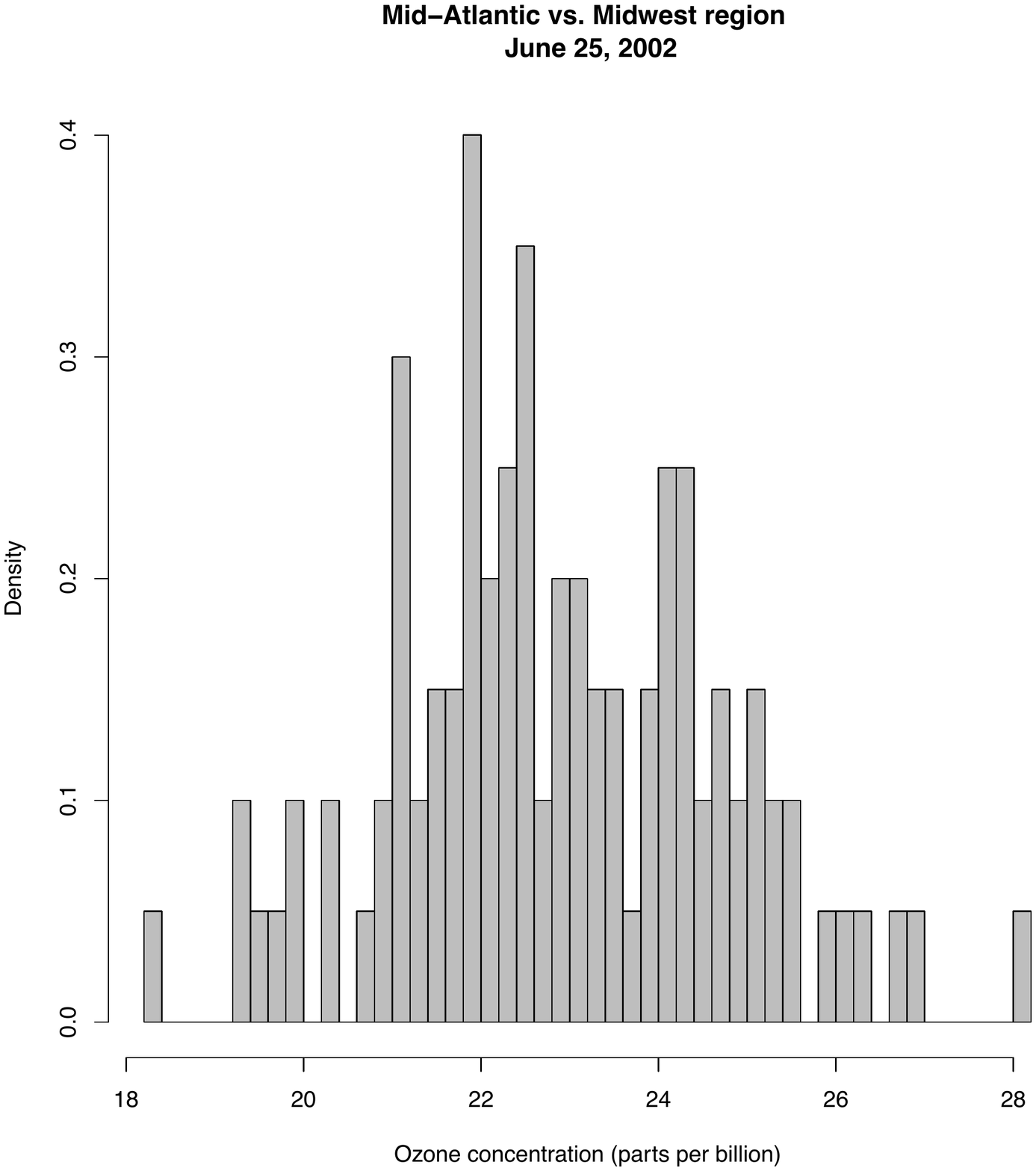}
&
\includegraphics[scale=0.2,angle=0]{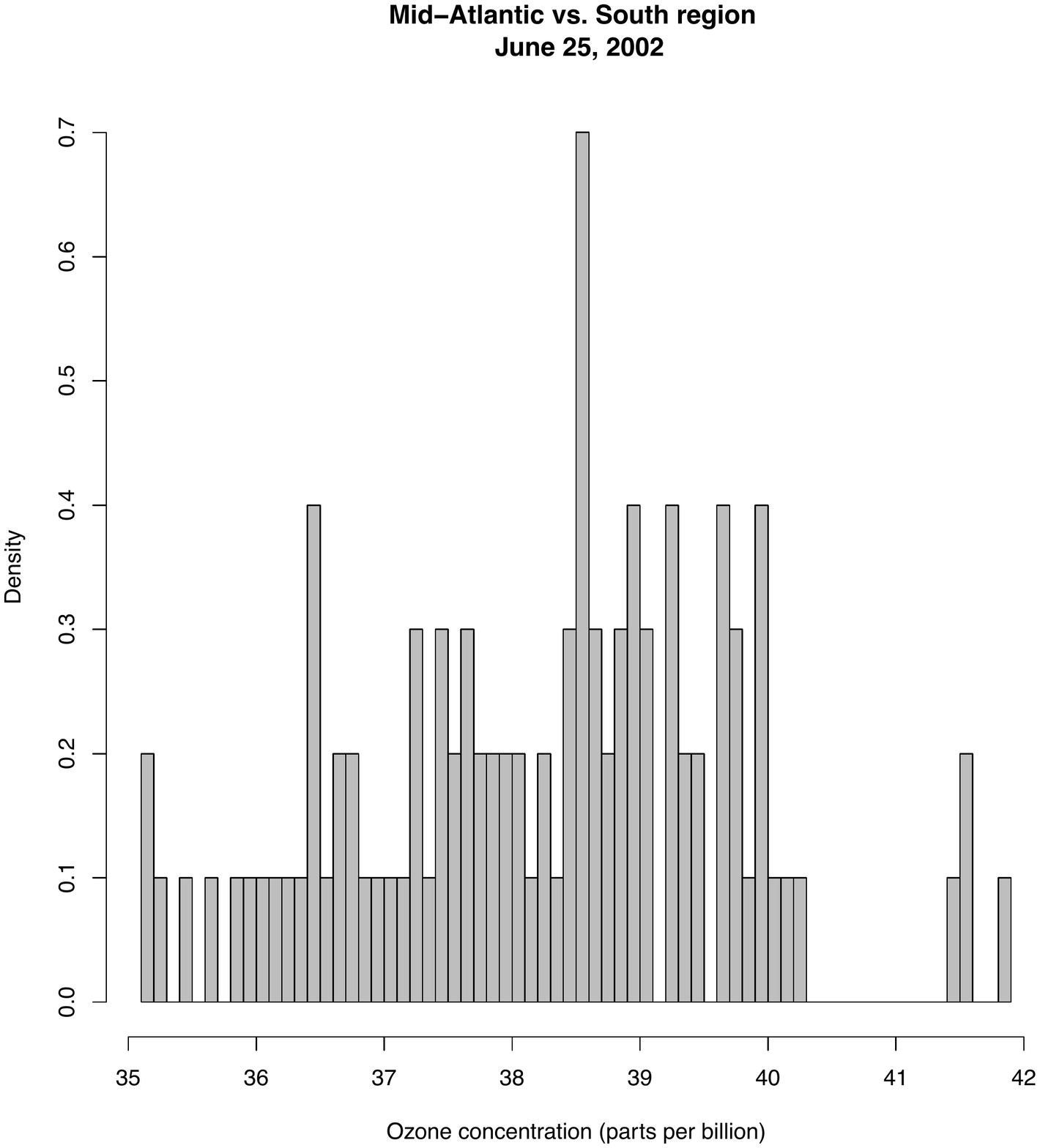}
&
\includegraphics[scale=0.2,angle=0]{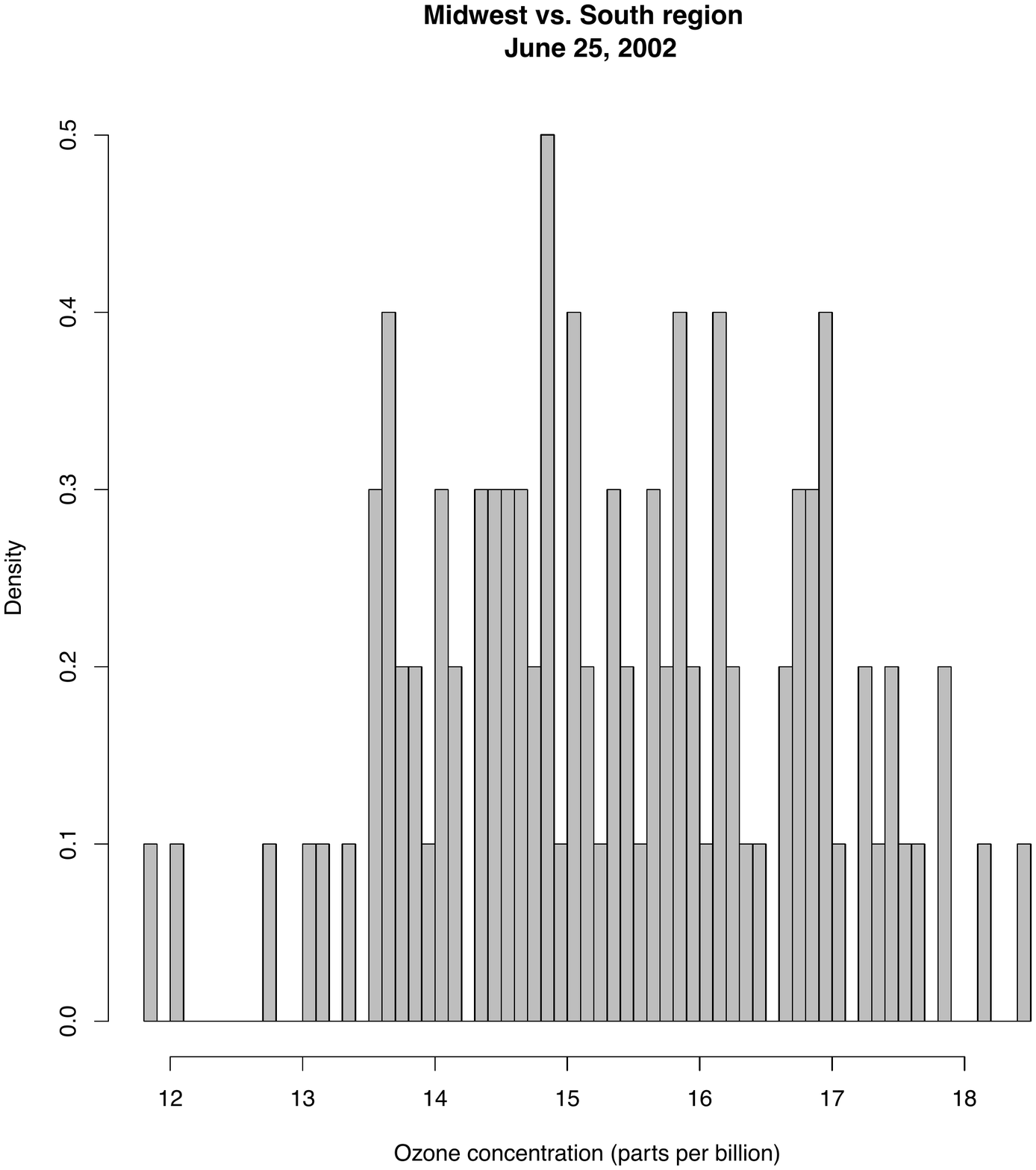}
\\
\includegraphics[scale=0.2,angle=0]{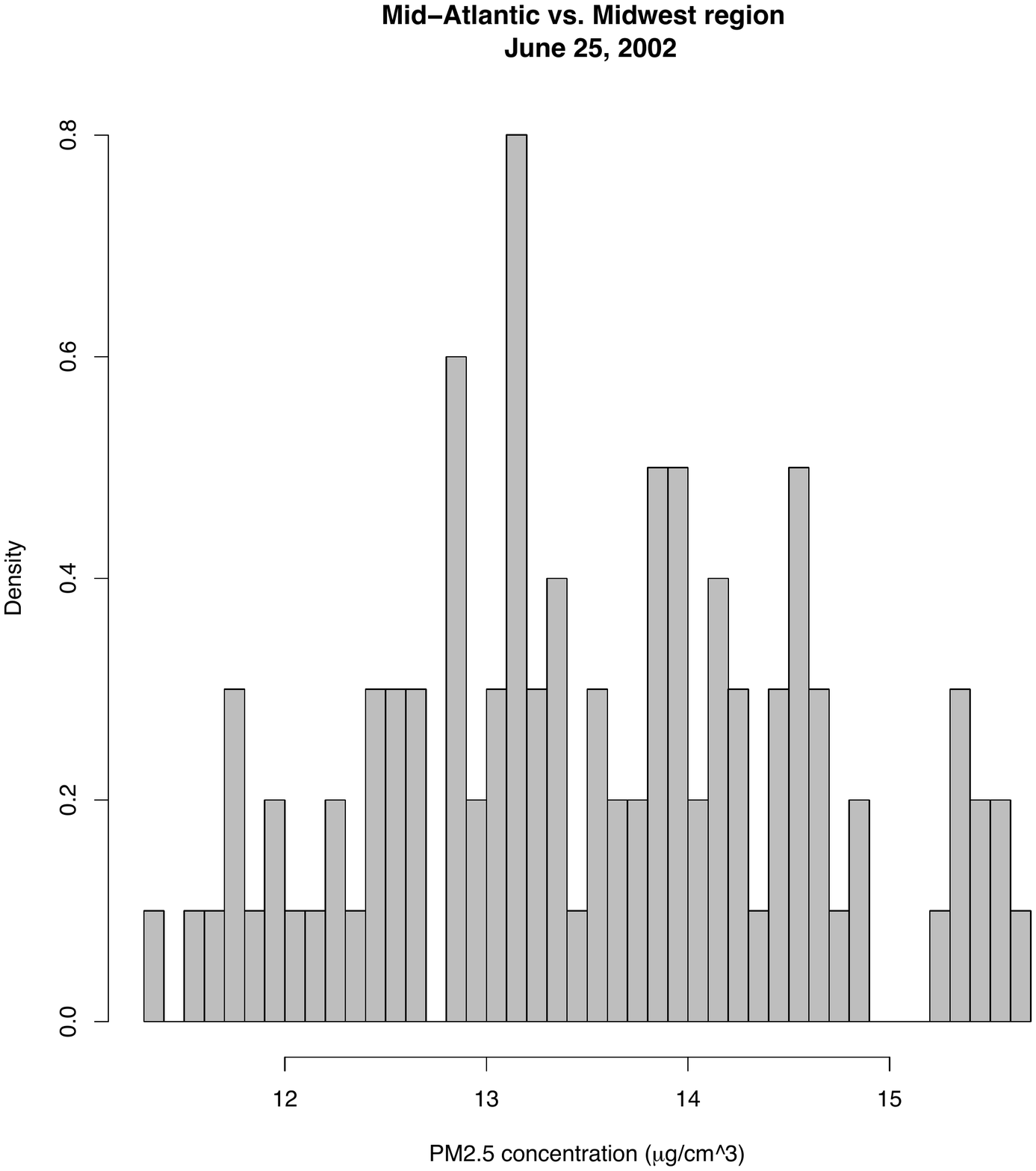}
&
\includegraphics[scale=0.2,angle=0]{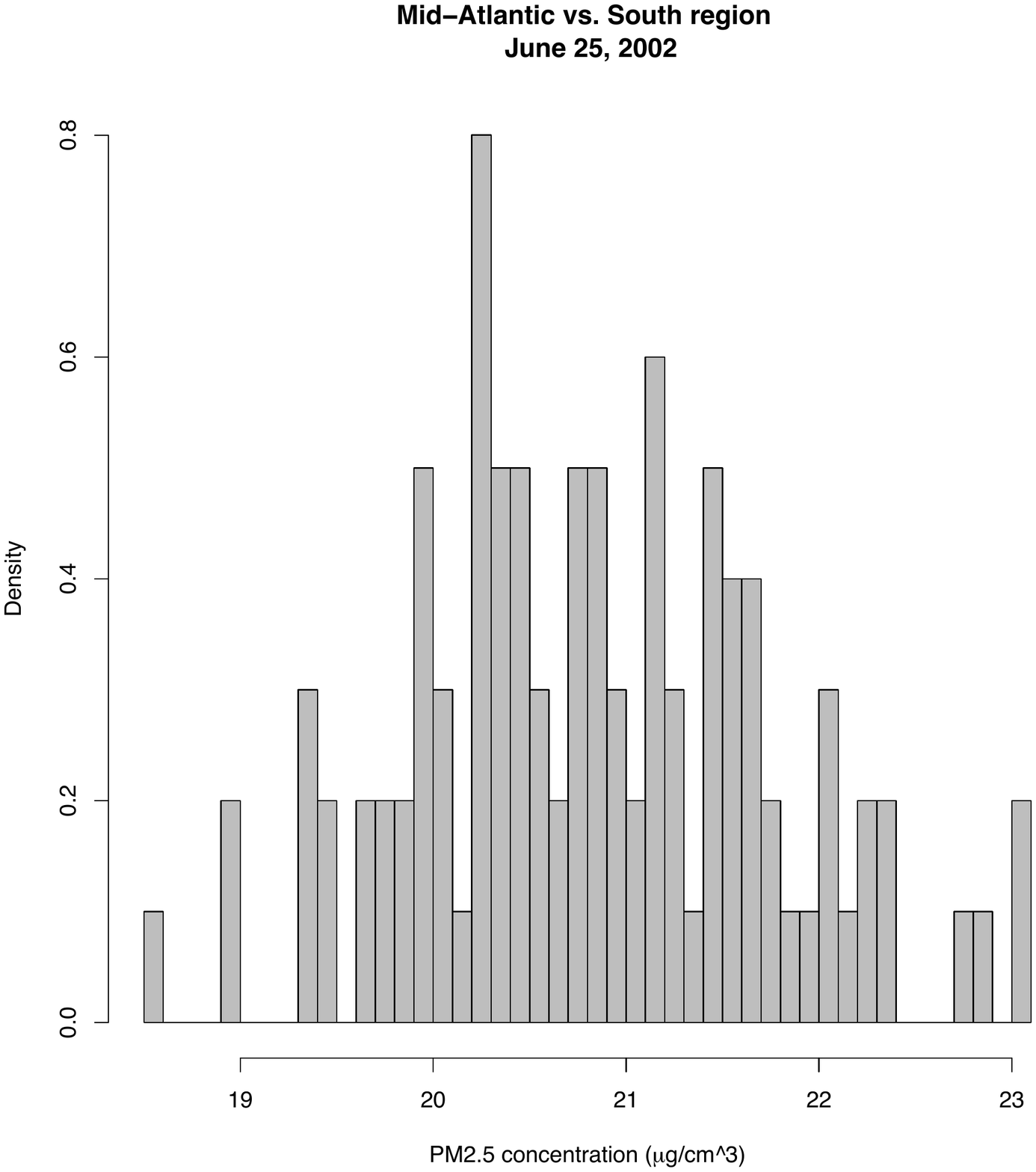}
&
\includegraphics[scale=0.2,angle=0]{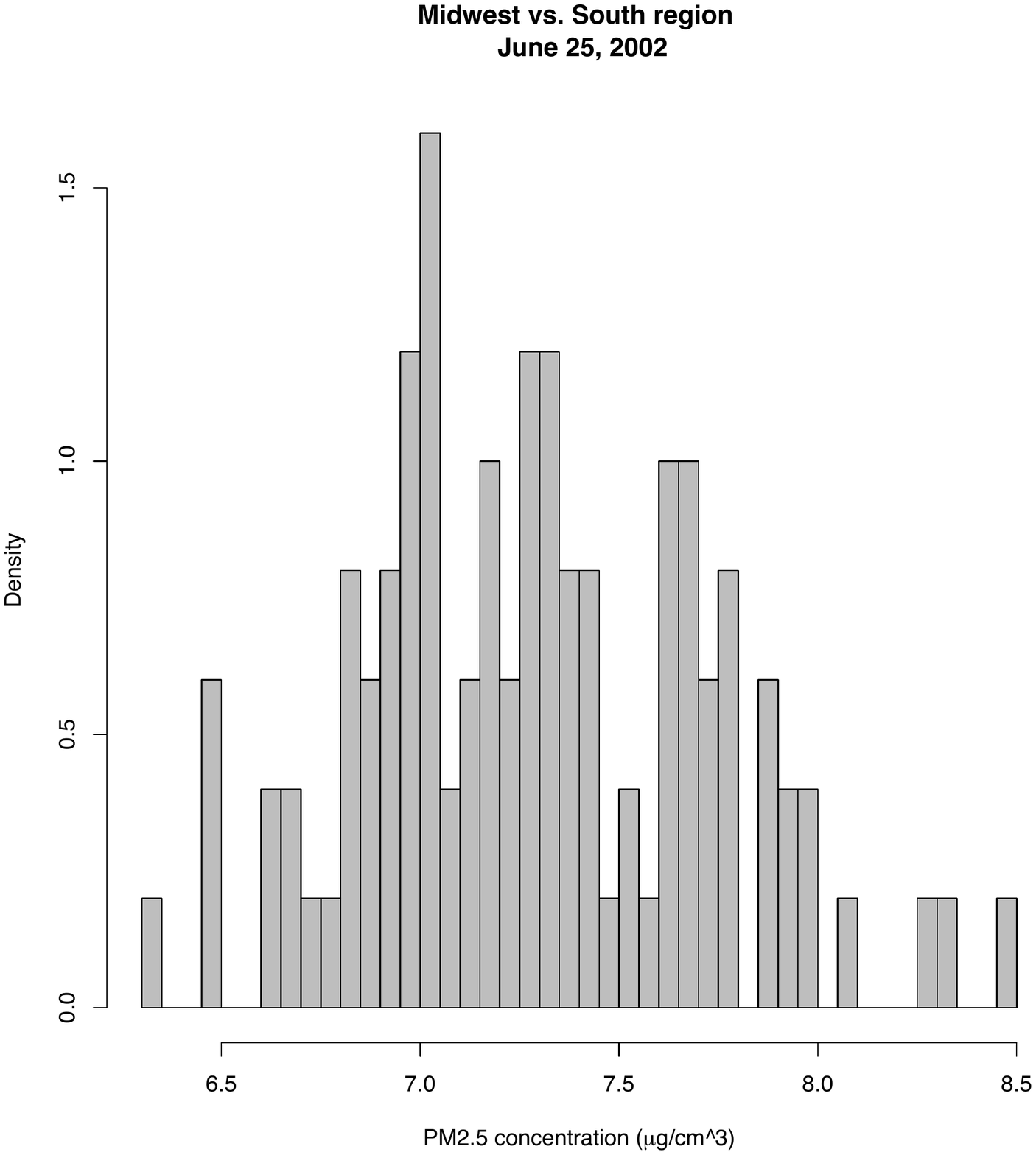}
\\
\end{tabular}
\vspace{0.2cm}
\caption{Posterior predictive distribution for the difference in average ozone and PM$_{2.5}$ concentration between Mid-Atlantic and
Midwest, Mid-Atlantic and South and Midwest and South on June 25, 2002, as obtained using the bivariate downscaler.
\label{fig:contrast}}
\end{figure}

An additional feature of our downscalers is the possibility to upscale and predict the variable of interest at any spatial scale. To illustrate such capability for our bivariate downscaler, we have divided our study region into three subregions: the Mid-Atlantic (Virginia, West Virginia, Maryland, Delaware, and Pennsylvania), the Midwest (Kentucky, Ohio, Illinois, Indiana, and Missouri), and the South (all the remaining states contained in the study region). For each of these regions, we have predicted the average ozone and PM$_{2.5}$ concentration on June 25, 2002, by predicting on a grid of points using the posterior predictive distribution induced by our bivariate downscaler model and by averaging the predictions over each of the three regions. Those predictions were then used to consider contrasts that might be
 of interest. Hence, Figure~\ref{fig:contrast} presents plots of the posterior predictive distribution of the contrast in average ozone and PM$_{2.5}$ concentration, respectively, between the Mid-Atlantic and the Midwest, Mid-Atlantic and the South and the Midwest and the South for June 25, 2002. As Figure~\ref{fig:contrast} shows, both the average ozone and PM$_{2.5}$ concentration are higher in the Mid-Atlantic region than in the Midwest and the South on June 25, 2002, with the South having, on average, the lowest average concentration for both pollutants on the selected day.  We could further push such comparisons to look at averages or contrasts over time.

\section{Summary}\label{sec:summary}

We have proposed a bivariate downscaler model that allows us to scale down the outputs of a numerical model from grid level to point level.
The model is rather general and flexible and can easily be extended to a multivariate downscaler
model for $p$ numerical model outputs.  For the bivariate case, we have examined several simplifications of the general bivariate downscaler model, each inducing a different type
of correlation structure among the two components of the bivariate random vector. We have also shown how the model can be extended to accommodate
data collected over both space and time.  Also, as a process model, by suitable block averaging, inference can be scaled up to provide average exposure over arbitrary regions.  In addition, we have demonstrated the computational advantages of scaling down, due to the relative sparsity of monitoring data compared with model output data, and the computational feasibility for our models with large space-time datasets.
We acknowledge that the small decay parameters (equivalently, large range parameters) relative to the size of our region, suggest that a linear variogram associated with a valid covariance function over a bounded region might be an alternative modeling path.  In turn, this suggests the possibility of an approach based entirely on contrasts, though, given that our interest is in spatially varying regressions on the model output, it remains unclear how to proceed if such a modeling choice was undertaken.
 
In our case study, we have applied the bivariate downscaler to downscale and predict ozone and PM$_{2.5}$ concentration during the summer months of year 2002; however it is clear
that our model can be used for any pair of pollutants. For example, it would be interesting to apply the bivariate downscaler model to
predict wet and dry sulfate deposition, embedding the bivariate downscaling approach in the modeling context proposed by \cite{Sahu&2009a}.
Another potentially interesting analysis would be to work with O$_{3}$ and CO.

There are two extensions we are currently exploring. First, in both the univariate and the bivariate downscaler model, the numerical model output
has been taken as data and we have not accounted for uncertainty in the predictions given by the numerical model.  Here, potential issues involve
too much smoothing of the CMAQ output, potential displacement of CMAQ grid cells, and errors in inputs that drive CMAQ.
Within our downscaling approach it is possible to characterize the uncertainty in the numerical model output by adding
another hierarchical level to our framework.  That is, we return to the issue in the Introduction where we now require a stochastic model for the CMAQ output.

Secondly, we seek extension beyond Gaussian specifications.  For instance, if we were to study extreme fields for two pollutants from both monitoring data and
CMAQ, we would not use Gaussian models.  Similarly, if we recorded two binary variables at each location, resulting in a $2 \times 2$ table for the location,
again, Gaussian processes are not appropriate.

\section{Appendix}\label{sec:appendix}
\subsection{Covariance and cross-covariance}\label{subsec:covariances}

Here we present explicit formulas for the covariance and cross-covariance between $Y_{1}(\mathbf{s})$ and $Y_{2}(\mathbf{s})$ for the different versions of the bivariate downscaler considered in Section
\ref{subsec:bivariate}.

For the bivariate downscaler with coregionalization matrix $\mathbf{A}$ with non-null entries, $A_{11}$, $A_{41}$, $A_{44}$, the covariance between the square root of the observed
ozone concentration, $Y_{1}(\mathbf{s})$ and the logarithm of the observed PM$_{2.5}$ concentration at $\mathbf{s}^{\prime}$ is given by (\ref{eq:corr_bidown_mod1}).
Furthermore, for this model:
\begin{eqnarray}
\mbox{Cov}(Y_{1}(\mathbf{s}),Y_{1}(\mathbf{s}^{\prime})) & = & A_{11}^{2} \exp(-\phi_{1} \| \mathbf{s} - \mathbf{s}^{\prime} \|)  +  \tau^{2}_{1} \delta_{\mathbf{s},\mathbf{s}^{\prime}} \nonumber \\
\mbox{Cov}(Y_{2}(\mathbf{s}),Y_{2}(\mathbf{s}^{\prime})) & = & A_{41}^{2} \exp(-\phi_{1} \| \mathbf{s} - \mathbf{s}^{\prime} \|) +
A_{44}^{2} \exp(-\phi_{4} \| \mathbf{s} - \mathbf{s}^{\prime} \|) +
\tau^{2}_{2} \delta_{\mathbf{s},\mathbf{s}^{\prime}}
\nonumber
\end{eqnarray}
where $\delta_{\mathbf{s},\mathbf{s}^{\prime}}$ is equal to 1 if $\mathbf{s} \equiv \mathbf{s}^{\prime}$ and equal to 0 otherwise.

For the bivariate downscaler with coregionalization matrix $\mathbf{A}$ with non-null entries, $\left\{ A_{11}, A_{22}, A_{33}, A_{41}, A_{44}, A_{55}, A_{66} \right\}$, the cross-covariance between
$Y_{1}(\mathbf{s})$ and $Y_{2}(\mathbf{s}^{\prime})$ is given by (\ref{eq:corr_bidown_mod1}), while the expressions for the within pollutant covariances are, respectively:
\begin{eqnarray}
\mbox{Cov}(Y_{1}(\mathbf{s}),Y_{1}(\mathbf{s}^{\prime})) & = & A_{11}^{2} \exp(-\phi_{1} \| \mathbf{s} - \mathbf{s}^{\prime} \|) \nonumber \\
& + & \sum_{k=2}^{3} \left[ A_{kk}^{2} \cdot
x_{k-1}(B) x_{k-1}(B^{\prime}) \cdot \exp(-\phi_{k} \| \mathbf{s} - \mathbf{s}^{\prime} \|) \right] +
\tau^{2}_{1} \delta_{\mathbf{s},\mathbf{s}^{\prime}} \nonumber \\
\mbox{Cov}(Y_{2}(\mathbf{s}),Y_{2}(\mathbf{s}^{\prime})) & = & A_{41}^{2} \exp(-\phi_{1} \| \mathbf{s} - \mathbf{s}^{\prime} \|) +
A_{44}^{2} \exp(-\phi_{4} \| \mathbf{s} - \mathbf{s}^{\prime} \|)   \nonumber \\
& + & \sum_{k=5}^{6} \left[ A_{kk}^{2} \cdot x_{(k-4)}(B) x_{(k-4)}(B^{\prime}) \cdot \exp(-\phi_{k} \| \mathbf{s} - \mathbf{s}^{\prime} \|) \right] +
\tau^{2}_{2} \delta_{\mathbf{s},\mathbf{s}^{\prime}}
\nonumber
\end{eqnarray}
with $B$ and $B^{\prime}$ grid cell containing, respectively, $\mathbf{s}$ and $\mathbf{s}^{\prime}$.

Finally, for the bivariate downscaler model with coregionalization matrix $\mathbf{A}$ given by (\ref{eq:bidown_coreg_mod3}), the covariance between the two components $Y_{1}(\mathbf{s})$ and $Y_{2}(\mathbf{s}^{\prime})$ of the bivariate random vector $\mathbf{Y}(\mathbf{s})$ is given by:
\begin{eqnarray}
\mbox{Cov}(Y_{1}(\mathbf{s}),Y_{2}(\mathbf{s}^{\prime})) & = & A_{11} A_{41} \cdot \exp(-\phi_{1} \| \mathbf{s} - \mathbf{s}^{\prime} \|) \nonumber \\ 
& + & \sum_{k=2}^{3} \left[ A_{kk} A_{(k+3) k} \cdot x_{k-1}(B) x_{k-1}(B^{\prime}) \cdot \exp(-\phi_{k} \| \mathbf{s} - \mathbf{s}^{\prime} \|) \right]
\end{eqnarray}
while the expression for the inter-pollutant covariances are given, respectively, by:
\begin{eqnarray}
\mbox{Cov}(Y_{1}(\mathbf{s}),Y_{1}(\mathbf{s}^{\prime}))  & = &  A_{11} \exp(-\phi_{1} \| \mathbf{s} - \mathbf{s}^{\prime} \|) \cdot \left[ A_{11} + A_{21} (x_{1}(B) + x_{1}(B^{\prime})) \right. \nonumber \\
& + & \left. A_{31} (x_{2}(B) + x_{2}(B^{\prime})) \right]  \nonumber \\
& + & \sum_{k=2}^{3} \left[ A_{kk}^{2} \cdot x_{k-1}(B) x_{k-1}(B^{\prime}) \cdot \exp(-\phi_{k} \| \mathbf{s} - \mathbf{s}^{\prime} \|) \right] + \tau^{2}_{1} \delta_{\mathbf{s},\mathbf{s}^{\prime}} \nonumber
\end{eqnarray}
and
\begin{eqnarray}
\mbox{Cov}(Y_{2}(\mathbf{s}),Y_{2}(\mathbf{s}^{\prime})) & = & A_{41}^{2} \exp(-\phi_{1} \| \mathbf{s} - \mathbf{s}^{\prime} \|) \nonumber \\
& + & \sum_{k=2}^{3} \left[ A_{(k+3)k}^{2} x_{k-1}(B) x_{k-1} (B^{\prime}) \cdot \exp(-\phi_{k}  \| \mathbf{s} - \mathbf{s}^{\prime} \|) \right]  \nonumber \\
& + &  A_{44} \cdot \exp(-\phi_{4} \| \mathbf{s} - \mathbf{s}^{\prime} \|) \cdot \left[ A_{44} + A_{54} (x_{1}(B) + x_{1}(B^{\prime})) \right. \nonumber \\
& + & \left. A_{64} (x_{2}(B) + x_{2}(B^{\prime})) \right]  \nonumber \\
& + & \sum_{k=5}^{6} A_{kk}^{2} \cdot x_{k-4}(B) x_{k-4}(B^{\prime}) \cdot \exp(-\phi_{k} \| \mathbf{s} - \mathbf{s}^{\prime} \|)  + \tau^{2}_{2} \delta_{\mathbf{s},\mathbf{s}^{\prime}} \nonumber
\end{eqnarray}

\vspace{0.3cm}
\noindent \textbf{Disclaimer:}
\newline The U.S. Environmental Protection Agency through its Office of Research and Development partially collaborated in the research described here. Although it has been reviewed by the Agency and approved for publication, it does not necessarily reflect the Agency's policies or views.

\bibliographystyle{imsart-number}
\bibliography{bidownscaler_paper_AOAS}

\end{document}